\documentclass[twoside,11pt]{article}

\usepackage{blindtext}

\usepackage{subcaption}
\usepackage{jmlr2e}
\usepackage{latexsym,amsfonts,tabularx,dsfont,mathrsfs}
\usepackage{bbm, mathtools, url}
\usepackage{stmaryrd}
\usepackage{enumerate}
\usepackage[table]{xcolor}
\usepackage{color}
\usepackage{algorithm, algorithmic}
\usepackage{multirow}
\usepackage{upgreek}
\usepackage{soul}
\usepackage{threeparttable}
\usepackage{booktabs}

\renewcommand{\dim}{d}

\newcommand{\inner}[2]{{#1}^T\!{#2}}

\newcommand{\reals}[1]{\mathbb R^{#1}}

\newcommand{\sphere}[1]{\mathbb S^{#1}}
\newcommand{\subsphere}[2]{{\mathbb{S}^{#1}_{#2}}}

\newcommand{\geodesic}[1]{\gamma_{#1}}
\newcommand{\tangentbundle}{\mathcal{S}\sphere{\dim -1}}
\newcommand{\gflow}[1]{\psi_{#1}}

\newcommand{\levelset}[1]{L(#1)}
\newcommand{\glevelset}[3]{L(#1,#2,#3)}

\newcommand{\area}[1]{\omega_{#1}}

\newcommand{\givens}[3]{R(#1, #2, #3)}
\renewcommand{\givens}[3]{G(#1, #2, #3)}

\newcommand{\vMF}{p_{\text{vMF}}}
\newcommand{\bing}{p_{\text{Bing}}}
\newcommand{\normal}[2]{\mathcal N(#1, #2)}
\newcommand{\uniform}[1]{\mathcal U_{(#1)}}
\newcommand{\unisphere}[1]{\sigma_{#1}}
\newcommand{\unitangsphere}[1]{\mu_{#1}}
\newcommand{\liouville}{\mathcal{L}}
\newcommand{\leb}{\lambda}
\newcommand{\shrinkagedist}[3]{Q_{{#1,}{#2,}{#3}}}
\newcommand{\shrink}[3]{\text{shrink}(#1,#2,#3)}

\newcommand{\sigmaalgebra}[1]{\mathcal{B}(#1 )}

\newcommand{\targetdensity}{p}
\newcommand{\target}{\pi}
\newcommand{\normconsttarget}{Z}

\renewcommand{\d}{{\rm d}}

\newcommand{\norm}[1]{\left\Vert #1 \right\Vert}
\newcommand{\dtv}[2]{d_{tv}\left(#1,#2\right)}

\newcommand{\targetpoint}{\pmb{q}}
\newcommand{\targetcloud}{Q}
\newcommand{\sourcepoint}{\pmb{p}}
\newcommand{\sourcecloud}{P}
\newcommand{\bbox}{B}
\newcommand{\rotation}{\pmb{R}}

\renewcommand{\algorithmicrequire}{\textbf{input:\ }}
\renewcommand{\algorithmicensure}{\textbf{output:\ }}

\usepackage[symbol]{footmisc} %makes footnotes as symbols instead of numbers
\usepackage{lastpage}
\jmlrheading{26}{2025}{1-\pageref{LastPage}}{9/23; Revised 10/25}{11/25}{23-1158}{Michael Habeck and Mareike Hasenpflug and Shantanu Kodgirwar and Daniel Rudolf}

\ShortHeadings{Geodesic Slice Sampling on the Sphere}{Habeck, Hasenpflug, Kodgirwar and Rudolf}
\firstpageno{1}

\begin{document}

\title{Geodesic Slice Sampling on the Sphere}
% Should we stick to this author ordering? I do not see a specific rule,
% if we want to highlight equally important authors, then we need to use,
% an asterisk
\author{\name Michael Habeck$^{1}$ \email michael.habeck@uni-jena.de
%	\addr Friedrich Schiller University Jena\\
%	07743 Jena, Germany
	\AND
	\name Mareike Hasenpflug$^{2}$ \email mareike.hasenpflug@uni-passau.de
%	\addr University of Passau\\
%	94032 Passau, Germany
	\AND
	\name Shantanu Kodgirwar$^{1}$ \email shantanu.kodgirwar@uni-jena.de
	\AND
	\name Daniel Rudolf$^{\,2}$ \email daniel.rudolf@uni-passau.de\\[1ex]
		\addr $^1$Friedrich Schiller University Jena,
	07743 Jena, Germany\\
	\addr $^2$Universit\"at Passau, Innstra\ss e 33, 94032 Passau, Germany}

% need to add the name of the editor eventually
\editor{Anthony Lee}

% renews the command by again allowing numerals on footnotes
\renewcommand{\thefootnote}{\arabic{footnote}}

\maketitle

\begin{abstract}%   <- trailing '%' for backward compatibility of .sty file
	Probability measures on the sphere form an important class of statistical models and are used, for example, in modeling directional data or shapes.
	Due to their widespread use, but also as an algorithmic building block, efficient sampling of distributions on the sphere is highly desirable.
	We propose a shrinkage based and an idealized geodesic slice sampling Markov chain, designed to generate approximate samples from distributions on the sphere.
	In particular, the shrinkage-based version of the algorithm can be implemented such that it runs efficiently and has no tuning parameters.
	We verify reversibility and prove that under weak regularity conditions geodesic slice sampling is uniformly 
	ergodic.
	Numerical experiments show that the proposed slice samplers achieve excellent mixing on challenging targets including distributions arising in rigid-registration problems
%	the Bingham distribution
    and mixtures of von Mises-Fisher distributions.
	In these settings our approach outperforms standard samplers such as random-walk Metropolis-Hastings and Hamiltonian Monte Carlo.
\end{abstract}


\section{Introduction}

%1. 	Introductiory sentences

In recent years, with the advent of sampling methods based on Markov chains, Bayesian inference with posterior distributions on manifolds attracted considerable attention.
In particular, various Markov chain algorithms for approximate sampling 
on different manifolds have been developed \citep[see][]{ByrneGirolami,Lan14,GoyalAndShetty,DimIndependentMCMCOnSpheres,beskos2022mcmc}.
Here we focus on the prototypical case of the sphere embedded in $\reals{\dim}$ with its inherent geometrical features
as the underlying manifold. Following the slice sampling paradigm, we introduce and analyze an efficient way of approximate sampling of distributions on the sphere.

%2.	Problem description

Let $\sphere{\dim -1}$ be the $(\dim-1)$-dimensional Euclidean unit sphere in $\reals{\dim}$ and let $\unisphere{\dim-1}$ denote the volume measure on $\sphere{\dim -1}$. For $\targetdensity\colon  \sphere{\dim -1} \to (0,\infty) $ satisfying
\begin{equation}\label{Eq: Integrability assumption on target density}
	\normconsttarget \coloneqq \int_{\sphere{\dim -1}} \targetdensity(x) \unisphere{\dim -1}(\d x) \in (0, \infty),
\end{equation}
we are interested in a target distribution $\target$ of the form
\begin{equation}\label{Eq: Definition of target}
	\target(\d x) = \frac{1}{\normconsttarget}\, \targetdensity(x) \unisphere{\dim -1}(\d x),
\end{equation}
such that $\targetdensity/Z$ is the probability density function of $\target$ relative to $\unisphere{\dim -1}$. In a Bayesian setting, $\target$ can be considered a posterior distribution determined by likelihood $p$ and prior measure $\unisphere{\dim-1}$.
Throughout the paper, we assume that $p$ can be evaluated, which is a common minimal requirement in the Markov chain Monte Carlo (MCMC) literature.
%3.	Motivation why interesting
Our motivation for considering MCMC on the sphere is twofold:\\[-3ex]
\begin{enumerate}
	\item \textit{Posteriors naturally defined on the sphere require efficient sampling:} To provide some examples,
	sampling distributions on the sphere plays an important role in directional statistics and shape analysis \citep[see][]{Mardia09}, and Bayesian inverse problems on $\sphere{2}$ occur, e.g., in astrophysics or geophysics \citep[see][]{Marignier}. Moreover, Bayesian density estimation \citep{holbrook2020nonparametric} requires sampling of a posterior on an infinite-dimensional sphere, which usually is approximated by truncating the dimension, ending up with $\sphere{\dim - 1}$ for large $\dim$.
	Also the problem of 3D rigid registration can be reduced to sampling from a distribution on $\sphere{3}$ (see Section \ref{Sec: rigid registration}).\\[-3.5ex]
	\item \textit{Spherical sampling can be used as tool within transforming Markov chains:}
	\citet{Lan14} introduced a Hamiltonian Monte Carlo (HMC) scheme for sampling from spherical distributions 
	that leads, after suitable transformations, to an efficient exploration of distributions in $\reals{\dim}$ constrained by inequalities.
 Moreover, recently \citet{StereographicProjection} demonstrated that MCMC algorithms on $\sphere{\dim-1}$ can target distributions in $\reals{\dim}$ by mapping them to the sphere by means of the stereographic projection and then sampling from the transformed target on the sphere. This is further developed by \cite{bell2024adaptive} with an adaptive MCMC strategy. The authors report that the ``stereographic scheme'' performs well in the presence of heavy-tailed distributions, as shown by both theoretical and empirical evidence. \citet{bell2024adaptive} also
indicate that the combination with slice sampling on the sphere, proposed here, is particularly effective and outperforms competing methods.\\ [-3ex]
% 	observe that in particular a combination 
% 	of slice sampling on the sphere (proposed in this paper) outperforms other approaches.
\end{enumerate}

%  4. 	Slice sampling paradigm -- > Algorithm

The advantage of the slice sampling paradigm is that it enables an automatic, location-dependent traversal of the space via density based regions, without requiring any laborious manual tuning of parameters to ensure efficient exploration of the state space.
	This usually results in an algorithm that is less sensitive to certain degenerations of the target distribution such as multimodality or anisotropy,
	see also
%	The advantage of following a slice sampling paradigm is that it allows for an exploration of the space that is inherently, i.e. without any tuning of parameters, adapted to the current location of the sampler.
%	The result is usually an algorithm that is less sensitive to certain degenerations of the target distribution such as multimodality or anisotropy,
 \cite{Neal,Murray}. A transition mechanism for realizing a corresponding Markov chain works as follows: Given current state $x\in\sphere{\dim-1}$, a superlevel set $L(t)$ of $\targetdensity$ is randomly determined by choosing a level $t\in (0, \targetdensity(x))$. Then, the next Markov chain instance is specified by (mimicking) sampling of the normalized reference measure $\sigma_{\dim-1}$ restricted to the level set. The latter step requires some care regarding its algorithmic design. 
%  5.  Main results
The main contributions of our paper are:\\[-3ex]
\begin{itemize}
	\item We introduce an ideal geodesic slice sampler with corresponding Markov kernel $H$
	for approximately simulating $\target$. 
	The kernel $H$ is implemented by moving along a randomly chosen great circle through the current state. The resulting sampling on the great circle (understood as geodesic) intersected with the level set can be algorithmically realized by using an univariate acceptance/rejection approach.\\[-4ex]
	\item To increase the computational efficiency 
		we propose a modification 
	of $H$
	by employing a shrinkage procedure (cf. \cite{Murray} and \cite{Neal})
	and call the resulting Markov kernel $\widetilde{H}$.
	\\[-4ex]
	\item
	We show that if $p$ is lower semicontinuous, i.e., the strict superlevel sets of $p$ are open, then both, $H$ and $\widetilde{H}$ are well-defined and reversible with respect to (w.r.t.) the target distribution $\target$. \\[-4ex]
	\item We prove uniform ergodicity for 
	lower semi\-con\-tinuous
	$\targetdensity$ that satisfy a  boundedness condition, i.e., we provide total variation distance estimates of the $n$th marginal of a Markov chain with transition kernel $H$ or $\widetilde{H}$ to $\target$, see Theorem \ref{Thm: Uniform ergodicity ISS} and Theorem \ref{Thm: Uniform geometric ergodicity of SSS}.\\[-4ex]
	\item We test our algorithms on challenging targets such as 
	a posterior distribution arising in rigid registration
%	{\textcolor{red}{the Bingham distribution}} 
	and mixtures of von Mises-Fisher distributions. We observe that, in a variety of settings, our slice samplers outperform standard samplers such as random walk Metropolis Hastings (RWMH) and Hamiltonian Monte Carlo (HMC) on spherical distributions.\\[-3ex]
\end{itemize}

% 7. 	Literature review
Let us comment on how our contributions fit into the literature. \citet{DimIndependentMCMCOnSpheres} present two MCMC algorithms to sample distributions on the sphere using push forward kernels of a preconditioned Crank-Nicolson algorithm and of elliptical slice sampling, respectively. The crucial difference to our work is that \citet{DimIndependentMCMCOnSpheres} assume that the target density is defined relative to an angular Gaussian distribution. Instead, here we consider target densities relative to the volume measure of the sphere. Observe that with increasing dimension both reference measures become ``increasingly singular" to each other. Furthermore, \citet{Marignier} propose a proximal MCMC method to sample from posterior distributions of inverse imaging problems on $\sphere{2}$. An infinite-dimensional setting of Bayesian density estimation is treated by \citet{holbrook2020nonparametric} by an HMC algorithm on a sphere. Other MCMC approaches have been developed on more general manifolds that usually cover $\sphere{\dim-1}$ as a special case. This includes the HMC algorithm for manifolds embedded in the Euclidean space introduced by \citet{ByrneGirolami} which uses the geodesic flow, 
and the HMC algorithm for manifolds that are the fibre of a smooth map proposed by \citet{brubaker2012family}.
Furthermore, \citet{GoyalAndShetty} investigate a Metropolis-Hastings-like geodesic walk on manifolds with non-negative curvature.
Similarly, the Metropolis-Hastings algorithm of \citet{zappa2018monte} based on projections from the tangent spaces can be applied to $\sphere{d-1}$.
Here it is also worth to mention that several geodesic walks exist that target the uniform distribution on subsets of the sphere. We refer to Section~\ref{Sec: Walking the sphere} for more details.
Finally, note that recently there has been some theoretical progress in the investigation of convergence properties of slice sampling \citep[see for example][]{GeometricConvergenceEllipticalSliceSampler,SimpleSliceSamplingSpectralGap,HybridSliceSampling,ReversibilityEllipticalSliceSampler} that very much influenced the presentation and proof arguments of our theoretical results.

%		8.	Outline
The outline of our paper is as follows: We start by introducing notation and details regarding the geometry of the sphere. In Section~\ref{Sec: Geodesic slice sampling}, we formulate the ideal and shrinkage slice sampler in terms of the transition kernel and transition mechanism. Here we also prove the reversibility and uniform ergodicity statements. Next we illustrate the applicability of our approach in different scenarios 
by numerical experiments. We conclude with a discussion. 

\section{Preliminaries}
In this section we provide some general notation, give a brief introduction to geometrical features of the Euclidean unit sphere
and explain how these can be leveraged to ``walk the sphere".

\subsection{Setting and Notation}
Let $\leb$ be the Lebesgue measure on $\reals{}$, and let $\norm{\cdot}$ be the Euclidean norm induced by the standard inner product $\inner{x}{y}$, where $x,y \in \reals{\dim}$ for $\dim \in \mathbb{N}$.
	Moreover, define $\mathrm{Id}_{\dim} \in \mathbb{R}^{\dim\times \dim}$ to be the $\dim$-dimensional identity matrix.
	Throughout this paper we assume $\dim \geq 3$ and consider the Euclidean unit sphere
	\[
	\sphere{\dim -1} \coloneqq \{ x \in \mathbb{R}^{\dim} \mid \norm{x} = 1\}
	\]
	 equipped with its Borel-$\sigma$-algebra $\sigmaalgebra{\sphere{\dim-1}}$. By $\unisphere{\dim-1}$ we denote the canonical volume measure on $(\sphere{\dim-1},\sigmaalgebra{\sphere{\dim-1}})$ that serves as reference measure for distributions, cf. \eqref{Eq: Definition of target}. To assess the difference of distributions $\nu$, $\mu$ on $(\sphere{\dim-1}, \sigmaalgebra{\sphere{\dim-1}})$ we use the total variation distance 
	 \[
	 \dtv{\mu}{\nu} \coloneqq \sup_{A \in \sigmaalgebra{\sphere{\dim-1}}} |\mu(A) - \nu(A)|.
	 \]
	 On a generic Borel space $(\mathcal{X}, \sigmaalgebra{\mathcal{X}})$ define the restriction of a measure $\nu$ to a set $A \in \sigmaalgebra{\mathcal{X}}$ by  $\nu\vert_A$, that is $\nu\vert_A(B) = \nu(A \cap B)$ for all $B \in \sigmaalgebra{\mathcal{X}}$. Whenever appropriate, for a set $A\in\sigmaalgebra{\mathcal{X}}$ we denote by $\mathcal{U}_A$ the uniform distribution on $A$. For example, for $\mathcal{X}=\mathbb{R}$ with $A\in\sigmaalgebra{\mathbb{R}}$ the probability measure $\mathcal{U}_A$ refers to $\frac{1}{\leb(A)} \, \leb\vert_A$ if $\leb(A)\in (0, \infty)$ and for $\mathcal{X}=\sphere{\dim-1}$ the distribution  $\mathcal{U}_{\sphere{\dim-1}}$ refers to $\frac{1}{\area{\dim -1}}\unisphere{\dim-1}$, where $\area{\dim -1}\coloneqq \unisphere{\dim-1} \left(\sphere{\dim -1}\right)$.
Sometimes it is more convenient to work with random variables. Therefore, let $(\Omega, \mathcal{F}, \mathbb{P})$ be a sufficiently rich probability space on which all random variables occurring in this paper are defined. If a random variable $X$ has distribution $\nu$, we write $X \sim \nu$.

\subsection{The Geometry of the Sphere}\label{S: Geodesics on the sphere}
To approximately sample from distributions on the sphere we rely on its (special) geometry.
The key objects in this context are great circles and ``equators''.
More formally regarding the latter, for  given $x \in \sphere{\dim -1}$ we call
\[
\subsphere{\dim - 2}{x} \coloneqq \{ v\in\sphere{\dim-1} \mid \inner{v}{x}=0 \}
\]
the great subsphere with pole $x$.
This $(\dim -2)$-dimensional subsphere is the intersection of $\sphere{\dim -1}$ and the $(\dim -1)$-dimensional hyperplane perpendicular to $x$, 
and in this sense can be thought of as a generalization of the equator.
We also equip $\subsphere{\dim -2}{x}$ with the corresponding volume measure, which we denote as $\unitangsphere{x}$.
Since $\subsphere{\dim -2}{x}$ is essentially ``a tilted $\sphere{\dim -2}$'',
we have $\unitangsphere{x}(\subsphere{\dim -2}{x}) = \area{ \dim -2}$  for all $x \in \sphere{ \dim -1}$, cf. Appendix~\ref{Sec: Sampling mu_x}.
Accordingly $\frac{1}{\area{\dim -2}}\unitangsphere{x}$ 
coincides with $\mathcal{U}_{\subsphere{\dim -2}{x}}$. In our transition mechanism, sampling w.r.t. the latter distribution frequently appears, such that we provide a procedure for performing it in Appendix~\ref{Sec: Sampling mu_x}. 

The great circles of the sphere are the ``largest'' $1$-dimensional subspheres spanning $\sphere{\dim - 1}$. 
Rigorously, for every pair $(x,v)$, where $x \in \sphere{\dim -1}$ and $v \in \subsphere{\dim -2}{x}$, we define the great circle
\[
	\geodesic{(x,v)}: \mathbb{R} \to \sphere{\dim -1}, \qquad \theta \mapsto \cos(\theta) x + \sin(\theta) v.
        \]
Intuitively, we can also think of $\geodesic{(x,v)}$ as the curve obtained by ``moving from the point $x$ in the direction of $v$ with a constant velocity''.
This interpretation originates from Riemannian geometry
and following its terminology we may use the terms great circle and geodesic interchangeably for the object $\geodesic{(x,v)}$.
%\replaced{we subsequently also use the term geodesic in the context of great circles.}{we use the terms great circle and geodesic interchangeably for the object $\geodesic{(x,v)}$.}
Note that all great circles passing through a point $x \in \sphere{\dim -1}$ are of the from $\geodesic{(x,v)}$ for some $v \in \subsphere{\dim -2}{x}$, 
i.e., the great circles through $x \in \sphere{\dim -1}$ are parametrized by $\subsphere{\dim -2}{x}$.

Observe that, due to the periodicity of sine and cosine,  all great circles are $2\uppi$-periodic. This is exploited when we incorporate great circles into our slice sampling approach.

We continue by illuminating an interaction between the geometric and the measure theoretic structure of the sphere with the help of the map
\begin{align*}
	T_\theta: \tangentbundle \to \tangentbundle,    \qquad
	(x,v)                    \mapsto \big(\cos(\theta)x +\sin(\theta)v,\, \sin(\theta) x - \cos(\theta) v\big)
\end{align*}
for $\theta \in \mathbb{R}$, 
where for brevity we define $\tangentbundle \coloneqq \bigcup_{x \in \sphere{\dim-1}} \big(\{x\} \times \subsphere{\dim-2}{x}\big)$.
If we interpret for a pair $(x,v) \in \tangentbundle$ the first component as position on the sphere and the second component as ``direction of view'',
then applying the map $T_\theta$ to $(x,v)$ can be thought of as ``following the great circle $\geodesic{(x,v)}$ from $x$ in direction $v$ for a length of $\theta$ and then doing a U-turn back on the spot''.
This intuition is underpinned by the following lemma, proven in Appendix \ref{Sec: Liouville measure invariant under T}, which essentially tells us that following the great circle corresponding to the pair $T_\theta(x,v)$ is the same as
following the original great circle $\geodesic{(x,v)}$ in reverse direction with an offset of $\theta$.

\begin{lemma}\label{L: T indentities geodesic}
	Let $x \in \sphere{\dim-1}$, $v \in \subsphere{\dim -2}{x}$. 
	For all $\theta, r \in \mathbb{R}$ we have $\geodesic{(x,v)}(\theta -r ) = \geodesic{T_\theta(x,v)}(r)$.
	In particular, this implies $\geodesic{T_\theta(x,v)}(\theta) = x$.
\end{lemma}
Moreover, central to our proof techniques is the observation that the measure on $\tangentbundle$ that ``zips up'' the volume measures $\unitangsphere{x}$ on the single fibers of $\tangentbundle$ through the volume measure $\unisphere{\dim -1}$ on $\sphere{\dim -1}$ is invariant under $T_\theta$.
\begin{lemma}\label{L: Liouville measure invariant under T}
	Let $F: \tangentbundle \to \mathbb{R}$ be a function such that 
	the left-hand side of the equation below exists. Then, for all $\theta \in \mathbb{R}$ we have
	\begin{equation}\label{Eq: Liouville measure invariant under T}
		\int_{ \sphere{\dim -1}} \int_{\subsphere{\dim -2}{x}} F(x,v) \unitangsphere{x}(\d v) \unisphere{\dim -1}(\d x)
		= \int_{ \sphere{\dim -1}} \int_{\subsphere{\dim -2}{x}} F\left(T_\theta(x,v)\right) \unitangsphere{x}(\d v) \unisphere{\dim -1}(\d x).
	\end{equation}
\end{lemma}
For the convenience of the reader we provide the proof of the former lemma in Appendix~\ref{Sec: Liouville measure invariant under T}. For identity \eqref{Eq: Liouville measure invariant under T} in a setting of more general manifolds we refer to \cite[Proof of Theorem 27]{GoyalAndShetty}.

\subsection{Geodesic Walk on the Sphere}\label{Sec: Walking the sphere}

	Based on great circles we may ``walk the sphere'' starting at some state $x\in\sphere{\dim-1}$ by iteratively performing the following:
\\[-3.5ex]
\begin{enumerate}
	\item Choose a great circle $\gamma$ through the given point $x$ randomly.
	\\[-4ex]
	\item Choose the next point on the great circle $\gamma$ randomly.
	\\[-3.5ex]
\end{enumerate}
For the first step in the outlined transition we can use the fact that great circles through $x \in \sphere{\dim -1}$ are parametrized by $\subsphere{\dim -2}{x}$. 
Namely, we realize ``a random great circle'' as $\gamma_{(x,v)}$, where $v$ is a sample from 
	$\mathcal{U}_{\subsphere{\dim -2}{x}}$, see Algorithm~\ref{alg:sample-subsphere} in Appendix~\ref{Sec: Sampling mu_x} for the simulation of this probability measure.
The distribution of the randomly chosen point on the great circle in step 2 significantly influences the stationary distribution of the random walk on the sphere. 
In this section, we consider it to be independent of $x$, which causes the stationary distribution to be the uniform one. In Section~\ref{Sec: Geodesic slice sampling}, we explain how the second step can be modified so as to obtain a Markov chain with a desired stationary probability measure.

Using a distribution $\tau$ on $[0, 2\uppi)$ to sample on the great circle, we obtain the following Markov kernel
\[
K_\tau(x,A) \coloneqq \int_{\subsphere{\dim -2}{x}} \int_{[0, 2\uppi)}\mathbbm{1}_A\left(\geodesic{(x,v)}(\theta) \right) \tau(\d \theta)\, \mathcal{U}_{\mathbb{S}_x^{d-2}}(\d v)
\]
with $x \in \sphere{\dim -1}, A  \in \sigmaalgebra{\sphere{\dim -1}}$. The corresponding transition mechanism is described in Algorithm~\ref{A: Geodesic random walk}, where
$\tau$ can be interpreted as the distribution of the step-size.
\begin{algorithm}[H]
	\caption{Geodesic random walk on the sphere.}\label{A: Geodesic random walk}
	\algorithmicrequire  current state $x \in \sphere{\dim-1}$\\
	\algorithmicensure next state $x'$
	\begin{algorithmic}[1]
%		\STATE Draw $V \sim \frac{1}{\area{\dim -2}}\,\unitangsphere{x}$, call the result $v$.
		\STATE Draw $V \sim \mathcal{U}_{\sphere{\dim-2}_x}$, call the result $v$. \qquad
 		\COMMENT{Perform Algorithm~\ref{alg:sample-subsphere} with input $x$.}
		\STATE Draw $\Theta \sim \tau$, call the result $\theta$.
		\STATE Set $x'= \cos(\theta) x + \sin(\theta) v$.
	\end{algorithmic}
\end{algorithm}
By Lemma~\ref{L: Liouville measure invariant under T} and Lemma~\ref{L: T indentities geodesic} we have for any $A, B \in \sigmaalgebra{\sphere{\dim -1}}$ that
\begin{align*}
	& \int_{B} K_\tau(x,A) \ \unisphere{\dim -1}(\d x) 
	 =  \int_{[0, 2\uppi)} \int_{\sphere{\dim -1}} \int_{\subsphere{\dim -2}{x}} \mathbbm{1}_B(x) \mathbbm{1}_A\big(\geodesic{(x,v)}(\theta)\big) \ \frac{\unitangsphere{x}(\d v)}{\area{\dim -2}}\, \unisphere{\dim-1}(\d x)\, \tau(\d \theta)\\
	 &= \int_{[0, 2\uppi)} \int_{\sphere{\dim -1}} \int_{\subsphere{\dim -2}{x}} \mathbbm{1}_B\big(\geodesic{(x,v)}(\theta)\big) \mathbbm{1}_A(x) \ \frac{\unitangsphere{x}(\d v)}{\area{\dim -2}}\, \unisphere{\dim-1}(\d x)\, \tau(\d \theta) 
	= \int_{A} K_\tau(x,B) \ \unisphere{\dim -1}(\d x),
		\allowdisplaybreaks
\end{align*}	
which yields that $K_\tau$ is reversible w.r.t. $\mathcal{U}_{\sphere{\dim -1}}$, see also \citet{LeeVempala} or \citet[Theorem 27]{GoyalAndShetty}.

Note that, depending on how $\tau$ is chosen, $K_\tau$ coincides with different random walks that have been discussed in the literature. We add some remarks related to this point:
\medskip

\textbf{Related work.}
	If the step-size is chosen to be constant, that is, $\tau = \delta_{\varepsilon}$ for $\delta_{\varepsilon}$ being the Dirac measure on $\mathbb{R}$ at $\varepsilon \in (0,2\uppi)$, then $K_{\delta_{\varepsilon}}$ coincides with the geodesic walk from \citet{MangoubiAndSmith} on the sphere. They provide dimension independent mixing time results for this walk on manifolds with bounded positive sectional curvature.

	For targeting the uniform distribution on the sphere, the geodesic random walk introduced by {\citet[Algorithm 1, for their $K = \sphere{\dim -1}$]{GoyalAndShetty}} corresponds to $K_\tau$ with $\tau$ being the distribution of $\varepsilon R$, where $R$ is a chi-distributed random variable with $\dim-1$ degrees of freedom and fixed $\varepsilon > 0$. For manifolds with non-negative sectional curvature and bounded Riemannian curvature tensor, \citet{GoyalAndShetty}
	provide mixing time results if their unfiltered walk is modified to target the uniform distribution of a strongly geodesically convex subset of the ambient manifold.

 Moreover, the kernel $K_{\delta_\varepsilon}$ is related to the retraction-based random walk introduced by \citet{RetractionWalk}, where the geodesics are replaced by retractions, i.e., second order approximations of the geodesics. \citet{RetractionWalk} show that for $\varepsilon \to 0$ this algorithm can be used to sample paths of the Brownian motion on a general manifold. Note, however, that this causes the stationary distribution of the resulting Markov chain to deviate from the uniform distribution.

  General transitions according to $K_\tau$ can also be interpreted as the repeated action of random rotations. The $\dim$-dimensional rotation matrix
	\begin{equation}\label{Eq: Givens rotation}
		\givens{y}{z}{\theta} =\,\, \mathrm{Id}_\dim + (\cos(\theta) - 1) (yy^T + zz^T) + \sin(\theta)\, (yz^T - zy^T)
	\end{equation}
	is the Givens rotation acting in the plane spanned by two  orthogonal directions $y, z\in \sphere{\dim-1}$ where $\theta$ is the rotation angle \citep[see][]{Givens58}.	In this view, a transition of $K_\tau$ to the next state $X'$ is achieved by drawing an axis $V \sim \mathcal{U}_{\subsphere{\dim-2}{x}}$ and an angle $\Theta \sim \tau$ to form a random Givens rotation that is applied to the current state $x$, i.e.,
		$X' = \givens{V}{x}{\Theta}\, x.$

	A suitable reformulation of the latter can be viewed as a generalization of Kac's random walk on the sphere \citep[see e.g.,][]{Kac56, Pillai17}. To perform a transition of Kac's walk, draw $1 \le I < J \le \dim$ and $\Theta \sim \uniform{0, 2\uppi}$ randomly to generate the next state $X'$ by rotating the current state $x$, i.e., $X' = \givens{e_I}{e_J}{\Theta}\, x$,
	where $\{e_i\}_{i=1}^\dim$ is the standard basis of $\reals{\dim}$.
	Kac's walk also approximately simulates the uniform distribution, but chooses the plane of rotation from a discrete set, whereas in the generalized version 
	the plane of rotation changes continuously and always contains the current state $x$. \citet{Pillai17} show optimal mixing time results for Kac's walk.

\section{Geodesic Slice Sampling in $\sphere{\dim-1}$}\label{Sec: Geodesic slice sampling}

Following the slice sampling paradigm, we construct Markov chains for approximate sampling of $\target$
 that rely on exploring the sphere along suitable 1-dimensional objects, the great circles.
For this purpose let 
\[
\levelset{t} := \{ x \in \sphere{\dim -1} \mid \targetdensity(x) > t \},\qquad t \in (0,\infty),
\]
be the superlevel set of $p$. Recall that $p$ specifies $\target$, cf. \eqref{Eq: Definition of target}.

The rough idea is to choose a great circle $\gamma_{(x,v)}$ and a level $t$ randomly and then, from the intersection of the corresponding superlevel set and the great circle, draw the next state suitably.
For $x \in \sphere{\dim-1}$, $v \in \subsphere{\dim -2}{x}$ and $t \in (0, \infty)$ we require
\[
\glevelset{x}{v}{t} := \{ \theta \in [0, 2\uppi) \mid \targetdensity( \geodesic{(x,v)}(\theta)) > t \},
\]
called
geodesic level set, 
which contains all points on the great circle $\geodesic{(x,v)}$ with the function value of $p$ being greater than $t$.
Naturally, this implies that level sets and geodesic level sets are linked via the identity
\begin{equation}\label{Eq: Level set identity}
	\mathbbm{1}_{\levelset{t}}\big( \geodesic{(x,v)}(\theta) \big) = \mathbbm{1}_{\glevelset{x}{v}{t}}(\theta).
\end{equation}
Exploiting lower semicontinuity of $p$ we are able to identify a regime, where the geodesic level sets have strictly positive Lebesgue measure.
\begin{lemma} \label{lem: low_semi_implies_geo_lev_>0}
	For $p$ being lower semicontinuous, any $x\in\sphere{\dim-1}$, $t\in (0,\targetdensity(x))$ and $v\in \subsphere{\dim -2}{x}$
	we have $\lambda(\glevelset{x}{v}{t})>0$. Moreover, the essential supremum
	norm $\norm{p}_\infty$ of $p$ w.r.t. $\sigma_{\dim-1}$ coincides with $\sup_{x\in\sphere{\dim-1}} p(x)$.
\end{lemma}
\begin{proof}
	By the lower semicontinuity $p^{-1}((t,\infty))=\levelset{t}$ is open, and by the fact that $x\in \levelset{t}$, it is also non-empty. Moreover, since $\gamma_{(x,v)}$ is continuous, we have that $\gamma_{(x,v)}^{-1}(\levelset{t})$ is again open and non-empty.
	Together with the $2\uppi$-periodicity of $\geodesic{(x,v)}$, this implies that $\glevelset{x}{v}{t} = \gamma_{(x,v)}^{-1}(\levelset{t}) \cap [0, 2\uppi)$ has non-empty interior, such that $\lambda(\glevelset{x}{v}{t})>0$. 	
	The final assertion follows by observing that $\levelset{\norm{p}_\infty}$ is open (by the lower semicontinuity) and by the definition of $\norm{p}_\infty$ satisfies $\sigma_{\dim-1}(L(\norm{p}_\infty))=0$. Consequently, we have $\levelset{\norm{p}_\infty}=\emptyset$, such that $\norm{p}_\infty = \sup_{x\in\sphere{\dim-1}}p(x)$.
\end{proof} 
In the following we always assume that $p$ is lower semicontinuous, which implies by the previous lemma the well-definedness of the slice sampling schemes that we introduce now.

\subsection{Ideal Geodesic Slice Sampling}

\begin{figure}[bt]
	\begin{subfigure}{0.3\textwidth}
		\includegraphics[width=\textwidth]{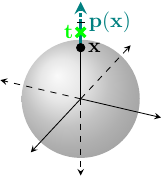}
		\caption{}
	\end{subfigure}
	\hfill
	\begin{subfigure}{0.3\textwidth}
		\includegraphics[width=\textwidth]{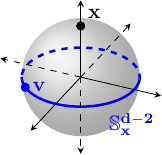}
		\caption{}
	\end{subfigure}
	\hfill
	\begin{subfigure}{0.3\textwidth}
		\includegraphics[width=\textwidth]{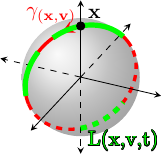}
		\caption{}
	\end{subfigure}
	\caption{Transition mechanism of the ideal geodesic slice sampler. 
		(\textit{a}) Sample a random level $t$ at the current point $x$.
		(\textit{b}) Sample a random point $v$ on $\subsphere{\dim -2}{x}.$
		(\textit{c}) Sample uniformly from the geodesic level set $\glevelset{x}{v}{t}$.}
	\label{F: illustration geoSSS-reject}
\end{figure}

We start by presenting an acceptance/rejection sampling based version of the geodesic slice sampler.
For $x \in \sphere{\dim -1}$, $t \in (0, \targetdensity(x))$ and $A \in \sigmaalgebra{\sphere{\dim -1}}$ let
\begin{align*}
	H_t(x,A) \coloneqq & \int _{\subsphere{\dim -2}{x}} \int_{\glevelset{x}{v}{t}} \mathbbm{1}_A( \geodesic{(x,v)}(\theta)) \, \frac{\d \theta}{\leb(\glevelset{x}{v}{t})} \, \frac{\unitangsphere{x}(\d v)}{\area{\dim -2}} \\
	= & \int _{\subsphere{\dim -2}{x}} \int_{\glevelset{x}{v}{t}} \mathbbm{1}_A ( \geodesic{(x,v)}(\theta)) \, \mathcal{U}_{\glevelset{x}{v}{t}} (\d \theta) \, \mathcal{U}_{\sphere{\dim-2}_x}(\d v).
\end{align*}
The kernel $H_t(x,\cdot)$ can be simulated by first choosing a random great circle $\geodesic{(x,v)}$,
and then, by sampling a point from the uniform distribution on $\glevelset{x}{v}{t}$, the intersection of the great circle and the level set $\levelset{t}$.
The \emph{ideal geodesic slice sampler} is given by the algorithm that implements a transition corresponding to the Markov kernel
\begin{equation}\label{Eq: Definition of H}
H(x,A) \coloneqq \frac{1}{\targetdensity(x)}\int_0^{\targetdensity(x)} H_t(x,A)\, \d t= \int_{0}^{p(x)} H_t(x,A)\; \mathcal{U}_{(0,p(x))}(\d t),
\end{equation}
where $ x \in \sphere{\dim -1}, A \in \sigmaalgebra{\sphere{\dim -1}}$. 
In words, a level $t\in (0,\targetdensity(x))$ is chosen uniformly distributed before $H_t(x,\cdot)$ is performed. 
A single transition of 
$H$
is described in Algorithm~\ref{A: ISS} 
and a graphical illustration on $\sphere{2}$ is provided in Figure \ref{F: illustration geoSSS-reject}.
\begin{algorithm}
	\caption{Ideal geodesic slice sampler.}\label{A: ISS}
	\algorithmicrequire  current state $x \in \sphere{\dim-1}$\\ %, $\dim  \geq 3$\\
	\algorithmicensure next state $x'$
	\begin{algorithmic}[1]
		\STATE Draw $V \sim \mathcal{U}_{\sphere{\dim-2}_x}$, call the result $v$. \qquad
		\COMMENT{Perform Algorithm~\ref{alg:sample-subsphere} with input $x$.}
		\STATE Draw $T \sim \uniform{0,\targetdensity(x)}$, call the result $t$. 
		\STATE \algorithmicrepeat
		\STATE \quad Draw $\Theta \sim \uniform{0, 2 \uppi}$, call the result $\theta$.
		\STATE \quad Set $x' = \cos( \theta) x + \sin(\theta) v$.
		\STATE \algorithmicuntil \ $\targetdensity(x')  > t$.
	\end{algorithmic}
\end{algorithm}

\begin{remark}\label{R: Relation to general framework}
	The geodesic slice sampler can be viewed as an instance of the procedure described in Section \ref{Sec: Walking the sphere} where the second step uses a distribution on the great circle that depends on the initial point $x$ and the chosen great circle $\gamma$.
\end{remark}
We establish the correctness of the ideal geodesic slice sampler by arguing for reversibility w.r.t. $\target$ and by proving a quantitative convergence guarantee.
The proof of the following result is provided in Appendix~\ref{sec: rev_iss}.	
\begin{proposition}
	\label{prop: rev_iss}
	\label{L: Reversibility of ISS}
	For $\target$ defined as in \eqref{Eq: Definition of target} with lower semicontinuous $p$, the kernel $H$ is reversible w.r.t. $\target$.
\end{proposition}

Now we present our main result for the ideal geodesic slice sampler: Imposing a boundedness assumption on $\targetdensity$, we provide an explicit convergence rate for the convergence to the stationary distribution in the total variation distance.
\begin{theorem}\label{Thm: Uniform ergodicity ISS}
For $p$ being lower semicontinuous with $\norm{p}_\infty<\infty$ we have
\begin{equation}
\label{eq: unif_erg}
	\sup _{x \in \sphere{\dim -1}} \dtv{H^n(x, \cdot	)}{\target} \leq \left( 1-\frac{\sup_{t>0} \left[t\cdot\mathcal{U}_{\sphere{\dim-1}}(L(t))\right]}{\sqrt{2\uppi} \sqrt{\dim-1}\norm{p}_\infty }\right)^n,
\qquad \forall n \in \mathbb{N},
\end{equation}
where $\sup_{t>0} \left[t\cdot\mathcal{U}_{\sphere{\dim-1}}(L(t))\right]\in (0,\norm{p}_\infty]$.
\end{theorem}

Let us comment on Theorem \ref{Thm: Uniform ergodicity ISS}.\\[-3ex]
\begin{remark}\label{R: r_p}
	The right-hand side of \eqref{eq: unif_erg} depends on $p$ through
	\[
		r_p:=\sup_{t>0} \left[t\cdot\mathcal{U}_{\sphere{\dim-1}}(L(t))\right]/\norm{p}_\infty.
		\]	
	 Observe that for lower semicontinuous $\targetdensity$ the size of $r_p \in (0,1]$ solely depends on the decreasing level set function $t\mapsto \mathcal{U}_{\sphere{\dim-1}}(L(t))$.
	 This function plays an even stronger role in controlling the convergence of ideal slice sampling. It is known to be the only relevant quantity for geometric ergodicity and spectral gap estimates, cf. \cite{roberts1999convergence,SimpleSliceSamplingSpectralGap}.
\end{remark}

\begin{remark} \label{R: expl_ex}
	The statement of the theorem contains the  fact that for $\norm{p}_\infty<\infty$ we have $r_p>0$, such that the right-hand side of \eqref{eq: unif_erg} converges exponentially to zero. We provide an example where $r_p$ is computed explicitly. For $\delta\in(0,1)$, $S\subseteq\sphere{\dim-1}$ open and $\mathcal{U}_{
	\sphere{\dim-1}}(S)=\gamma \in (0,1]$  consider $p(x)= \mathbbm{1}_{S}(x)+\delta\cdot \mathbbm{1}_{\sphere{\dim-1}\setminus S}(x)$. Note that for the corresponding level sets we have $\mathcal{U}_{\sphere{\dim-1}}(L(t))=\mathbbm{1}_{(0,\delta)}(t)+\gamma\cdot\mathbbm{1}_{[\delta,1)}(t)$ which implies  $r_p=\max\{\gamma,\delta\}$.
\end{remark}

\begin{remark} \label{R: har}
	We can interpret the ideal geodesic slice sampler as hit-and-run on the sphere. The hit-and-run algorithm has been introduced  for approximate sampling of distributions on $\mathbb{R}^\dim$ \citep[see][]{belisle1993hit}. A transition of the algorithm that starts in $x \in \mathbb{R}^\dim$ is performed in the following way: A direction $v$ is chosen uniformly at random in the unit ball centered around $x$ to construct a straight line $\ell$ through $x$ in direction $v$.	Then the next point is chosen w.r.t. the distribution of interest restricted to the line $\ell$. Since on the sphere straight lines correspond to great circles, ideal geodesic slice sampling can be considered as hit-and-run where simulating the target distribution restricted to the great circle is performed by slice sampling. Investigations on the performance of the hit-and-run algorithm in $\mathbb{R}^{\dim}$ w.r.t. the dimension attracted a lot of attention, cf. 
	\cite{lovasz2007geometry}. Regarding this aspect note that for $\varepsilon\in (0,1)$ we require 
	$n\geq \sqrt{2(d-1)\uppi}\, r_p^{-1} \log(\varepsilon^{-1})$ 
for achieving $\sup _{x \in \sphere{\dim -1}} \dtv{H^n(x, \cdot	)}{ \target} <\varepsilon$, which illuminates a moderate dependence on $d$.\\[-3ex]
\end{remark}

In the proof of Theorem~\ref{Thm: Uniform ergodicity ISS} we apply \citet[Theorem 16.2.4]{MeynTweedie}. For the convenience of the reader we provide a reformulation of the relevant parts of this result.

\begin{lemma}[{\citealp[Theorem 16.2.4]{MeynTweedie}}]\label{L: Meyn and Tweedie result}
	Let $P: \sphere{\dim -1} \times \sigmaalgebra{\sphere{\dim -1}} \to [0, 1]$ be a Markov kernel with stationary probability measure $\eta$. Assume there exists a non-zero measure $\nu$ on $\big( \sphere{\dim -1}, \sigmaalgebra{\sphere{\dim -1}} \big)$ such that
	$
	P(x,A) \geq \nu(A)
	$
	for all $x \in \sphere{\dim -1}$ and all $A \in \sigmaalgebra{\sphere{\dim -1}}$. Then 
	\[
	\sup_{x \in \sphere{\dim -1}} \dtv{P^n(x, \cdot)}{\eta } \leq \big(1 - \nu(\sphere{\dim -1})\big)^n, \qquad \forall n \in \mathbb{N}.
	\]
\end{lemma}

We add one more auxiliary result which is proven in Appendix~\ref{Sec: Integral estimate for smallness}.

\begin{lemma}\label{L: Essential step for smallness of S^d}
 For $x\in\sphere{\dim-1}$ and $A\in \sigmaalgebra{\sphere{\dim-1}}$ we have
 \[
 	\int_0^\uppi \int_{\subsphere{\dim -2}{x}} \mathbbm{1}_A\left(\geodesic{(x,v)}(\theta)\right)\, \mathcal{U}_{\subsphere{\dim-2}{x}}(\d v)\, \d \theta \geq 
 	\frac{\sqrt{2\uppi}}{\sqrt{d-1}}\;
 	\mathcal{U}_{\sphere{\dim-1}}(A).
 \]
\end{lemma}

Now we turn to proof of the theorem:

\begin{proof}\emph{(Proof of Theorem \ref{Thm: Uniform ergodicity ISS}.)}
	%%%%%%%%%%%%%%%%%%%%%%%%%%
	Let  $x \in \sphere{\dim -1}$ and $s\in(0,\norm{p}_\infty)$. Observe that for all $v \in \subsphere{\dim -2}{x}$ and $\theta \in [0, 2\uppi)$ with $\geodesic{(x,v)}(\theta) \in L(s)$ holds
	\begin{align} \label{al: trick_slava}
	\frac{1}{\targetdensity(x)} \int_0^{\targetdensity(x)} \mathbbm{1}_{\left[0, \targetdensity\left( \geodesic{(x,v)}(\theta) \right)\right)}(t) \, \d t
	= \min \left \{1, \frac{\targetdensity\left( \geodesic{(x,v)}(\theta) \right)}{\targetdensity(x)} \right\}
	\geq \frac{s}{\norm{p}_\infty}.
	\end{align}
	Observe that $\glevelset{x}{v}{t} \subseteq [0, 2 \uppi)$ such that $\leb(\glevelset{x}{v}{t}) \leq 2 \uppi$ for all $v \in \subsphere{\dim -2}{x}$ and all $t \in (0, \infty)$.	
	Then, using Lemma \ref{L: Essential step for smallness of S^d} this implies
	\begin{align*}
		 H(x,A)   
		 &= \frac{1}{\targetdensity(x)}\int_0^{\targetdensity(x)} \int _{\subsphere{\dim -2}{x}} \frac{1}{\leb(\glevelset{x}{v}{t})}
		 \int_{\glevelset{x}{v}{t}}\mathbbm{1}_A\big(\geodesic{(x,v)}(\theta)\big)
		 \, \d \theta\, \mathcal{U}_{\subsphere{\dim-2}{x}}(\d v) \, \d t                                                                                                                       \\  
		 &\geq \frac{1}{\targetdensity(x)}\int_0^{\targetdensity(x)} \int _{\subsphere{\dim -2}{x}} \frac{1}{\leb(\glevelset{x}{v}{t})}
		 \int_{\glevelset{x}{v}{t}}\mathbbm{1}_{A\cap L(s)}\big(\geodesic{(x,v)}(\theta)\big)
		 \, \d \theta\, \mathcal{U}_{\subsphere{\dim-2}{x}}(\d v) \, \d t                                                                                                                       \\  
		 &\geq \frac{1}{2\uppi} \int _{\subsphere{\dim -2}{x}} \int_0^{2\uppi} \mathbbm{1}_{A\cap L(s)}\big(\geodesic{(x,v)}(\theta)\big)
		 	\frac{1}{\targetdensity(x)} \int_0^{\targetdensity(x)} \mathbbm{1}_{\left[0, \targetdensity\left( \geodesic{(x,v)}(\theta) \right)\right)}(t) \, \d t
		 		 \, \d \theta\, \mathcal{U}_{\subsphere{\dim-2}{x}}(\d v)                                                                                                                      \\
		& \underset{\eqref{al: trick_slava}}{\geq} \frac{s}{2\uppi\norm{p}_\infty}  \int_0^{\uppi} \int _{\subsphere{\dim -2}{x}} \mathbbm{1}_{A\cap L(s)}\big(\geodesic{(x,v)}(\theta)\big)
		\,\mathcal{U}_{\subsphere{\dim-2}{x}}(\d v)                                    \, \d \theta
		 \geq \frac{s\; \mathcal{U}_{\sphere{\dim-1}}(A\cap L(s))}{\sqrt{2\uppi}\sqrt{d-1} \norm{p}_\infty} \,,
	\end{align*}
	for all $A \in \sigmaalgebra{\sphere{\dim-1}}$.
	By Lemma \ref{L: Meyn and Tweedie result} this implies
	\[
	\sup_{x \in \sphere{\dim -1}} \dtv{H^n(x, \cdot)}{\target} \leq \left( 1 - \frac{s \; \mathcal{U}_{\sphere{\dim-1}}(L(s))}{\sqrt{2 \uppi}\, \sqrt{\dim -1} \norm{p}_\infty}\right)^n
	\] 
	for any $s\in (0, \|p\|_\infty)$. For $s \geq \|p\|_\infty$ the statement holds trivially. Therefore taking the infimum over $s>0$ on the right hand-side gives \eqref{eq: unif_erg}. By the lower semicontinuity of $p$ we have that $L(s)$ is open and non-empty, such that $\mathcal{U}_{\sphere{\dim-1}}(L(s))\in (0,1]$ and consequently
	$0<s\;\mathcal{U}_{\sphere{\dim-1}}(L(s))<\norm{p}_\infty$ for all $s\in(0,\norm{p}_\infty)$, which provides the final claim.
\end{proof}
\subsection{Geodesic Shrinkage Slice Sampling}
Now we modify the ideal geo\-desic slice sampler by replacing the acceptance/rejection step of Algorithm~\ref{A: ISS} by
a shrinkage procedure that has been also used in elliptical slice sampling \citep[see][]{Murray} and originates in
Neal’s bracketing procedure on the interval $[0, 2\uppi)$ \citep[see][]{Neal}. 
We pick candidates from subsets of $[0, 2\uppi)$ that shrink until an accepted sample is generated. Intuitively, this strategy reduces the number of rejections per iteration, because candidates will be drawn from nested intervals that contain a neighborhood of the current state. 
Note that this approach can still be viewed as a realization of the procedure discussed in Section~\ref{Sec: Walking the sphere} in the sense of Remark \ref{R: Relation to general framework}.

\begin{figure}[bt]
	\begin{subfigure}[t]{0.24\textwidth}
		\includegraphics[width=\textwidth]{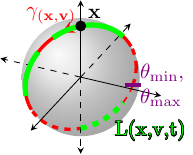}
		\caption{}
	\end{subfigure}
	\begin{subfigure}[t]{0.24\textwidth}
		\includegraphics[width=\textwidth]{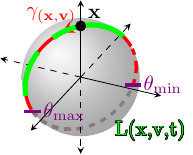}
		\caption{}
	\end{subfigure}
	\begin{subfigure}[t]{0.24\textwidth}
		\includegraphics[width=\textwidth]{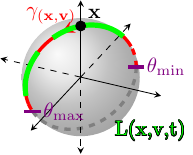}
		\caption{}
	\end{subfigure}
	\begin{subfigure}[t]{0.24\textwidth}
		\includegraphics[width=\textwidth]{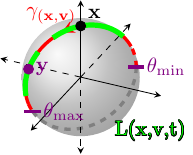}
		\caption{}
	\end{subfigure}
	\caption{Shrinkage procedure of the geodesic shrinkage slice sampler.
	(\textit{a}) First proposal is rejected.
	(\textit{b}) Second proposal is rejected and becomes new right bound of the proposal interval.
	(\textit{c}) Third proposal is rejected and becomes new left bound of the proposal interval.
	(\textit{d}) Fourth proposal lies in geodesic level set $\glevelset{x}{v}{t}$ and is accepted.}
	\label{F: illustration geoSSS-shrink}
\end{figure}

In Algorithm~\ref{A: Shrinkage procedure specific} we provide the shrinkage procedure as subroutine that is called as $\shrink{x}{v}{t}$ with input $x \in \sphere{\dim -1}$, $v \in \subsphere{\dim -2}{x}$ and level $t \in (0, \targetdensity(x))$. Roughly, it generates a point from a specified geodesic level set $\glevelset{x}{v}{t}$
by drawing points uniformly from a segment of (one winding) of the great circle $\geodesic{(x,v)}$ until we hit $\glevelset{x}{v}{t}$,
while the rejected points are used to successively shrink the segment of the great circle.
We work with this procedure as a black box by encapsulating it in the distribution of its output.
\begin{algorithm}
	\caption{Shrinkage procedure, called as $\shrink{x}{v}{t}$.}\label{A: Shrinkage procedure specific}
	\algorithmicrequire  current state $x \in \sphere{\dim -1}$, direction $v \in \subsphere{\dim -2}{x}$, level $t \in (0, \targetdensity(x))$\\
	\algorithmicensure parameter $\theta \in \glevelset{x}{v}{t}$
	\begin{algorithmic}[1]
		\STATE Draw $\Theta \sim \uniform{0,2\uppi}$, call the result $\theta$. 
		\STATE Set $\theta_{\min}=  \theta - 2 \uppi$ and set $\theta_{\max} = \theta$.
%		\STATE Set $\theta_{\max} = \theta$.
		\WHILE{$\targetdensity\left( \cos(\theta)x + \sin(\theta)v \right) \leq t$}
		\IF{$\theta < 0$}
		\STATE Set $\theta_{\min} =  \theta$.
		\ELSE
		\STATE Set $\theta_{\max} = \theta$.
		\ENDIF
		\STATE Draw $\Xi \sim \mathcal{U}_{(\theta_{\min}, \theta_{\max})}$, call the result $\theta$.
		\ENDWHILE
	\end{algorithmic}
\end{algorithm}
Considering the subroutine of Algorithm~\ref{A: Shrinkage procedure specific} as random variable yields the following definition.
\begin{definition}\label{D: Definition of shrinkage distribution}
	For lower semi-continuous $\targetdensity$, for $x \in \sphere{\dim -1}$, $v \in \subsphere{x}{\dim -2}$ and  $t \in (0, \targetdensity (x))$ we set
	\[
	\shrinkagedist{x}{v}{t}(A) := \mathbb{P}\big((\shrink{x}{v}{t} \mod 2\uppi)\in A\big), \qquad\forall A \in \sigmaalgebra{[0, 2\uppi)},
	\]
	where $\shrink{x}{v}{t}$ is determined by Algorithm~\ref{A: Shrinkage procedure specific}.
\end{definition}
\begin{remark}\label{R: Lower semi-continuity}
	The distribution $\shrinkagedist{x}{v}{t}$ coincides with the kernel of the shrinkage procedure defined by \citet[{Algorithm 2.2 with $\theta_{\text{in}}=0$, $S = \glevelset{x}{v}{t}$}]{ReversibilityEllipticalSliceSampler}. For details we refer to Appendix~\ref{Sec: Reversibility of SSS}. 
	The lower semicontinuity of $p$ ensures that $\shrinkagedist{x}{v}{t}$ is well-defined, cf. \citet{ReversibilityEllipticalSliceSampler}. Intuitively, the openness of the superlevel sets guarantees	that $x$ lies in a subinterval of the image of the geodesic level set $\geodesic{(x.v)}\big(\glevelset{x}{v}{t}\big)$ for all $v \in \subsphere{\dim -2}{x}$ and all $t \in (0, \targetdensity(x))$. Hence, while shrinking the sampling region on $\geodesic{(x,v)}([0, 2\uppi))$, the probability to hit the level set for a given level $t \in (0, \targetdensity(x))$ is strictly positive.	
\end{remark}
A single transition of the \emph{geodesic shrinkage slice sampler} is 
presented
in Algorithm~\ref{A: SSS}  
and for an illustration of the shrinkage scheme on a great circle on $\sphere{2}$ we refer to Figure \ref{F: illustration geoSSS-shrink}.
\begin{algorithm}
	\caption{Geodesic shrinkage slice sampler.}\label{A: SSS}
	\algorithmicrequire  current state $x \in \sphere{\dim-1}$\\
	\algorithmicensure next state $x'$
	\begin{algorithmic}[1]
		\STATE Draw $V \sim \mathcal{U}_{\sphere{\dim-2}_x}$, call the result $v$. \qquad		 		\COMMENT{Perform Algorithm~\ref{alg:sample-subsphere} with input $x$.}
		\STATE Draw $T \sim \uniform{0,\targetdensity(x)}$, call the result $t$. 
		\STATE Draw $\Theta \sim \shrinkagedist{x}{v}{t}$, call the result $\theta$. \qquad
		 		\COMMENT{Perform Algorithm~\ref{A: Shrinkage procedure specific} with input $x, v, t$.}
		\STATE Set $x' = \cos(\theta)x + \sin(\theta)v$.
	\end{algorithmic}
\end{algorithm}
To define the corresponding kernel $\widetilde{H}$, we need the following auxiliary level set kernels
$\widetilde{H}_t$.
Those correspond to drawing a random great circle through the starting point $x$,
and then running the shrinkage procedure to generate a point from the intersection of the random great circle and the level set $\levelset{t}$. 
That is, for $x \in \sphere{\dim-1}$, $t \in (0, \targetdensity(x))$ and $A \in \sigmaalgebra{\sphere{\dim -1}}$ we set
\begin{align*}
	\widetilde{H}_t(x,A)
	= \int _{\subsphere{\dim -2}{x}} \int_{[0,2\uppi)} \mathbbm{1}_A\big(\geodesic{(x,v)}(\theta) \big)
	\, \shrinkagedist{x}{v}{t}(\d \theta)\, \mathcal{U}_{\sphere{\dim-2}_x}(\d v)
\end{align*}
and let
\[
\widetilde{H}(x,A) = \frac{1}{\targetdensity(x)} \int_0^{\targetdensity(x)} \widetilde{H}_t(x,A) \,\d t
= \int_0^{\targetdensity(x)} \widetilde{H}_t(x,A) \, \mathcal{U}_{(0,p(x))}(\d t).
\]
In words, the kernel $\widetilde{H}$ can be described as first sampling a random level $t$ below the current value of the unnormalized density $\targetdensity$ and then running the auxiliary kernel $\widetilde{H}_t$ associated to that random level.
\begin{remark}
	Observe that due to the $2 \uppi$-periodicity of the great circle on the sphere, we have 
	\begin{align*} 
		&\cos\big(\shrink{x}{v}{t} \mod 2\uppi\big)x + \sin\big(\shrink{x}{v}{t} \mod 2\uppi\big)v\\
		&\qquad =  	\cos\big(\shrink{x}{v}{t} \big)x + \sin\big(\shrink{x}{v}{t} \big)v
	\end{align*}
	for all $x \in \sphere{\dim -1}$, $v \in \subsphere{x}{\dim -2}$ and  $t \in (0, \targetdensity (x))$. Therefore, we may introduce this modulo operation when transitioning from  the algorithmic formulation to a Markov kernel.\\[-4ex]
\end{remark}

In Appendix~\ref{Sec: Reversibility of SSS} we prove that $\widetilde{H}$ is reversible with respect to our target distribution, which implies that $\target$ is the stationary
distribution of this kernel. We obtain the following statement.\\[-4ex]
\begin{proposition}\label{L: Reversibility SSS}
	For $\target$ defined as in \eqref{Eq: Definition of target} with lower semicontinuous $p$, the kernel $\widetilde{H}$ is reversible w.r.t. $\target$.
	\\[-3ex]
\end{proposition}

For the geodesic shrinkage slice sampler we also provide a statement about convergence to the target distribution.

\begin{theorem}\label{Thm: Uniform geometric ergodicity of SSS}
	For $p$ being lower semicontinuous with $\norm{p}_\infty<\infty$ we have
	\begin{equation}
	\label{eq: unif_erg_SSS}
	\sup _{x \in \sphere{\dim -1}} \dtv{\widetilde{H}^n(x, \cdot	)}{\target} \leq \left( 1-\frac{\sup_{t>0} \left[t\cdot\mathcal{U}_{\sphere{\dim-1}}(L(t))\right]}{\sqrt{2\uppi} \sqrt{\dim-1}\norm{p}_\infty }\right)^n,
	\qquad \forall n \in \mathbb{N},
	\end{equation}
	where $\sup_{t>0} \left[t\cdot\mathcal{U}_{\sphere{\dim-1}}(L(t))\right]\in (0,\norm{p}_\infty]$.
\end{theorem}

 	    We follow the same strategy of proof as in Theorem~\ref{Thm: Uniform ergodicity ISS}, i.e.,
 		we show that the whole state space is a small set for $\widetilde{H}$ and then apply Lemma~\ref{L: Meyn and Tweedie result}. To this end, the crucial idea is to estimate the distribution $\shrinkagedist{x}{v}{t}$ by its restriction to the event that the loop in Algorithm~\ref{A: Shrinkage procedure specific} terminates after the first iteration. 
 	
\begin{proof}\emph{(Proof of Theorem~\ref{Thm: Uniform geometric ergodicity of SSS}.)}
		For arbitrary  $x \in \sphere{\dim -1}$, $s\in(0,\norm{\targetdensity}_\infty)$, for any $v \in \subsphere{\dim -2}{x}$ and $\theta \in [0, 2\uppi)$ with $\geodesic{(x,v)}(\theta) \in L(s)$ we have (cf. Equation \ref{al: trick_slava}) that
		\begin{equation}
	\label{al: trick_slava2}
	\frac{1}{\targetdensity(x)} \int_0^{\targetdensity(x)} \mathbbm{1}_{\left[0, \targetdensity\left( \geodesic{(x,v)}(\theta) \right)\right)}(t) \, \d t
	\geq \frac{s}{\norm{\targetdensity}_\infty}.
	\end{equation}
	Moreover, for all $t \in (0, \targetdensity(x))$ we obtain with $\Xi \sim \uniform{0, 2\uppi}$ that
	\begin{equation}
	\label{eq: est_Q}
	\shrinkagedist{x}{v}{t}(B) \geq \mathbb{P}\big(\Xi \in B \cap \glevelset{x}{v}{t}\big) = \frac{1}{2\uppi} {\leb}	\big(B \cap \glevelset{x}{v}{t}\big), \quad B \in \sigmaalgebra{[0, 2\uppi)}.
	\end{equation}
	Then
	\begin{align*}
%	\label{Eq: Lower estimate for SSS kernel}
		\widetilde{H}(x,A)
		& \underset{\eqref{eq: est_Q}}{\geq} \frac{1}{\targetdensity(x)} \int_0^{\targetdensity(x)}\int_{\subsphere{\dim -2}{x}} \frac{1}{2\uppi} \int_{\glevelset{x}{v}{t}} \mathbbm{1}_{A}\big(\geodesic{(x,v)}(\theta)\big) \, \d \theta \, \mathcal{U}_{\subsphere{\dim-2}{x}}(\d v)\, \d t\\
		& \geq \frac{1}{2\uppi} \int_{\subsphere{\dim -2}{x}} \int_{[0,2\uppi)} \mathbbm{1}_{A\cap L(s)}(\geodesic{(x,v)}(\theta)) \frac{1}{\targetdensity(x)}\int_0^{\targetdensity(x)}	\mathbbm{1}_{[0,\targetdensity(\geodesic{(x,v)}(\theta)))}(t) \d t\, \d \theta \,\mathcal{U}_{\subsphere{\dim-2}{x}}(\d v)\\
		& \underset{\eqref{al: trick_slava2}}{\geq} \frac{s}{2\uppi \norm{\targetdensity}_\infty}
			\int_{\subsphere{\dim -2}{x}} \int_{[0,2\uppi)} \mathbbm{1}_{A\cap L(s)}(\geodesic{(x,v)}(\theta)) \,\d \theta \,\mathcal{U}_{\subsphere{\dim-2}{x}}(\d v)
		\geq  \frac{s\cdot \mathcal{U}_{\sphere{\dim-1}}(A\cap L(s))}{\sqrt{2\uppi}{\sqrt{\dim-1}} \norm{\targetdensity}_\infty},
	\end{align*}
	where the last inequality follows by Lemma~\ref{L: Essential step for smallness of S^d}.
	Therefore, by Lemma \ref{L: Meyn and Tweedie result} and by taking the infimum over $s>0$ on the right hand-side we get
	\[
	\sup_{x \in \sphere{\dim -1}} \dtv{\widetilde{H}^n(x ,\cdot)}{\target} \leq
\left( 1-\frac{\sup_{t>0} \left[t\cdot\mathcal{U}_{\sphere{\dim-1}}(L(t))\right]}{\sqrt{2\uppi} \sqrt{\dim-1}\norm{p}_\infty }\right)^n, \quad \forall n\in\mathbb{N}.
	\]
	Finally, note that  $\sup_{t>0} \left[t\cdot\mathcal{U}_{\sphere{\dim-1}}(L(t))\right]\in (0,\norm{p}_\infty]$ follows already from Theorem~\ref{Thm: Uniform ergodicity ISS}. 
\end{proof}

 We comment on the result.
	\begin{remark}
			Note that 
			the right hand-side of \eqref{eq: unif_erg_SSS} coincides with the 
			the right hand-side of \eqref{eq: unif_erg} stated in Theorem~\ref{Thm: Uniform ergodicity ISS}, that addressed the uniform ergodicity statement for ideal slice sampling. This is due to the proof technique, since in the small set estimate, we lower bound both kernels with the same expression appearing in Lemma~\ref{L: Essential step for smallness of S^d}.
			Intuitively, it is clear that 
			$\sup_{x \in \sphere{\dim -1}} d_{tv}(H^n(x ,\cdot),\target)$			
			is smaller than 
			$\sup_{x \in \sphere{\dim -1}} d_{tv}(\widetilde{H}^n(x ,\cdot),\target)$, since the shrinkage procedure just adaptively imitates the acceptance/rejection step of the ideal one to gain computational efficiency. Exactly this gain in efficiency leads in applications to a potentially better accuracy to cost ratio, although the performance per Markov chain transition may be worse. 
			We also point out that 	
			 Remark~\ref{R: r_p} and Remark~\ref{R: expl_ex}
			apply to Theorem~\ref{Thm: Uniform geometric ergodicity of SSS} as well.
			\\[-4ex]
	\end{remark}

%  Daniel: The following remark is correct, but it does not add much. I think actually this is commonly known and for the sake of shortening the paper I am in favor of commenting it out.

%{\color{blue}
%We provide a final comment concerning the applicability of both geodesic slice sampling algorithms.
%\begin{remark}
%	Although we state the ideal geodesic slice sampler and the geodesic shrinkage slice sampler for strictly positive $\targetdensity$,
%	we can also treat unnormalized density functions that are $[0, \infty)$-valued using the following modification: 
%	In case $\targetdensity$ reaches zero, 
%	we restrict the state space of our slice sampling algorithms to the set $\{x \in \sphere{\dim -1} \mid \targetdensity(x) > 0\}$.
%	All arguments can then be performed analogously.
%\end{remark}
%}

\section{Numerical Illustrations}\label{Sec: Numerical experiments}
We apply geodesic slice sampling on the sphere for approximate sampling on various distributions on the sphere and compare it to tailor-made and general purpose MCMC algorithms including random-walk Metropolis-Hastings (RWMH) and  Hamiltonian Monte Carlo (HMC). The RWMH algorithm on the sphere is very similar to the standard RWMH, for details refer to  Appendix~\ref{app:rwmh}. \citet{Lan14} have developed a modified version of HMC for sampling spherical distributions. The details of the HMC algorithm can be found in Appendix \ref{app:hmc}. Both RWMH and HMC have a step-size parameter that is automatically tuned for each target, see Appendix \ref{Sec: Step-size tuning}. For a prespecified sampler we denote by $(x_n)_{n\in\mathbb{N}} \subset \sphere{d-1}$ a realization of the  corresponding Markov chain.

\subsection{Rigid Registration of Biomolecular Structures}\label{Sec: rigid registration}
First, we discuss an application of the geodesic slice sampler on the sphere to a common inference problem in structural bioinformatics, computer vision, and robotics. Two 3D objects are represented by point clouds, $\targetcloud=\{\targetpoint_i\}_{i=1}^I$ and $\sourcecloud=\{\sourcepoint_j\}_{j=1}^J$, where $\targetpoint_i, \sourcepoint_j \in \reals{3}$. The task is to find a 3D rotation matrix $\rotation\in SO(3)$ that best superimposes both point clouds. This is a rigid registration or pose estimation problem. In general, the point clouds have different size, and the correspondence between points in $\targetcloud$ and $\sourcecloud$ is unknown. 

We use a probabilistic model inspired by the coherent point drift (CPD) method \citep{Myronenko2010} to address this task. According to the model, a point in $\targetcloud$ is either generated from a Gaussian mixture model whose components have equal weights and are centered on the rotated source $\sourcecloud$, or it is an outlier that has no corresponding point in $\sourcecloud$. Outliers are generated from a uniform distribution $\mathcal{U}_\bbox$ over the bounding box $\bbox$ of the target point cloud (i.e. $\bbox\subset\reals{3}$ is the smallest rectangle such that $\targetcloud\subset\bbox$). We denote the probability density function w.r.t. the Lebesgue measure of this distribution as $p_o$. The probability that a point is an outlier is $\omega\in[0, 1]$. The probabilistic model, i.e., the conditional density again w.r.t. the Lebesgue measure, for a single point in $\targetcloud$ is 
\begin{equation}\label{eq:cpd}
  p(\targetpoint_i\mid \sourcecloud, \rotation, \omega, \sigma) = \omega p_o(\targetpoint_i) + \frac{1-\omega}{J\, (2\uppi\sigma^2)^{3/2}} \sum_{j=1}^J \exp\left(- \frac{1}{2\sigma^2} \|\targetpoint_i - \rotation\sourcepoint_j\|^2\right)\, .
\end{equation}
We can represent rotation matrices via unit quaternions \citep{Horn87} and thereby map the rigid registration problem to a distribution over $\sphere{3}$. This mapping is achieved by representing rotation matrices with unit quaternions:
\begin{equation}\label{eq:rotation-from-quaternions}
\rotation(x) = \begin{pmatrix}
  1 - 2 (x_3^2 + x_4^2) & 2 (x_2 x_3 - x_1 x_4) & 2 (x_2 x_4 + x_1 x_3) \\
  2 (x_2 x_3 + x_1 x_4) & 1 - 2 (x_2^2 + x_4^2) & 2 (x_3 x_4 - x_1 x_2) \\
  2 (x_2 x_4 - x_1 x_3) & 2 (x_3 x_4 + x_1 x_2) & 1 - 2 (x_2^2 + x_3^2)
\end{pmatrix}
\end{equation}
where $x=(x_1, x_2, x_3, x_4)^T \in \sphere{3}$. Therefore, probabilistic pose estimation boils down to generating samples from a distribution with density $p$ over $\sphere{3}$, where $x\in\sphere{3}$ encodes a 3D rotation matrix. The density $p$ w.r.t the volume measure on $\sphere{3}$ of the overall posterior is (up to normalization) the product of the densities in \eqref{eq:cpd}, that is,
\begin{equation}\label{eq:posterior-cpd}
  p(x) \propto \prod_{i=1}^I p(\targetpoint_i \mid \sourcecloud, \rotation(x), \omega, \sigma)\, .
\end{equation}

\begin{figure}[tb]
  \centering\includegraphics[width=\textwidth]{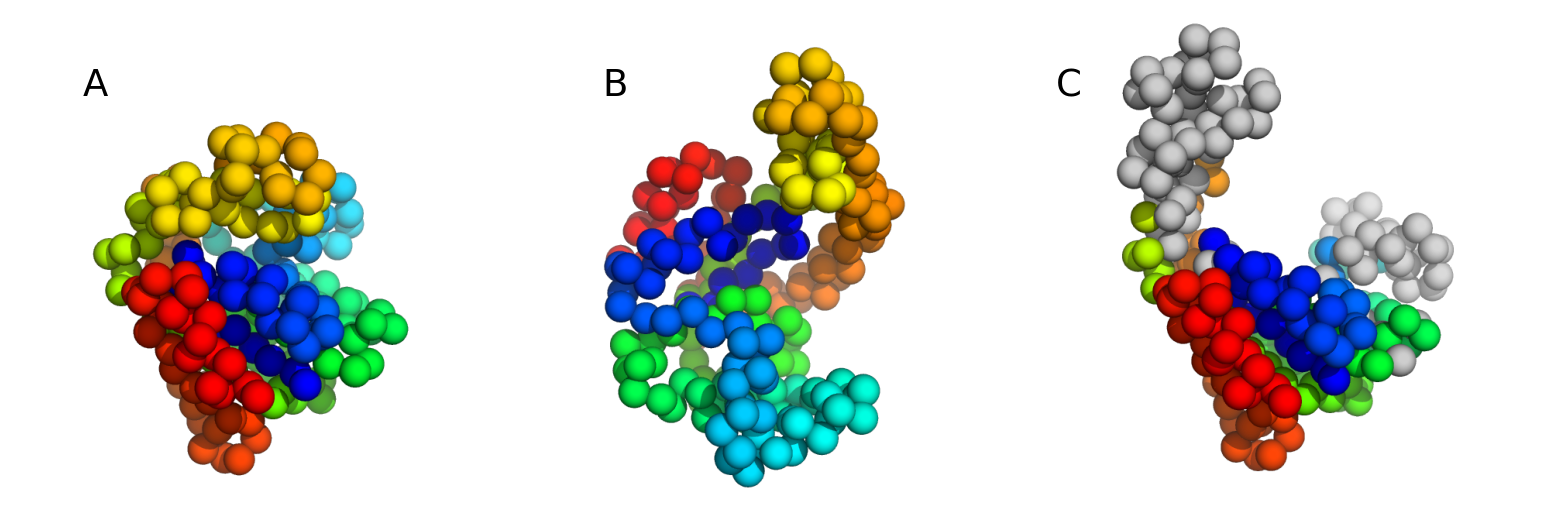}
  \caption{Rigid registration of point clouds. Each point cloud consists of 214 carbon-$\alpha$ atom positions. Points are colored according to their sequence position: N-terminal amino acids are shown in blue, C-terminal points are red. (A) Target point cloud showing AK in closed configuration. (B) Source point cloud showing AK in open configuration and in an orientation that  differs from the orientation of the target. Note that target and source points with identical color correspond to each other. However, due to the non-rigid conformational change not all source points will generate target points. (C) Source rotated optimally so as to generate the target with high probability according to the CPD model (\ref{eq:cpd}). Note that the points in the source that do not generate a target point are shown in gray. These are located in the moveable domains of the AK structure (for example the orange domain that is closed in the target, but open in the source).}\label{fig:proteins}
\end{figure}

In the following, we assume that the fraction of outliers $\omega$ and the standard deviation $\sigma$ are known such that the only unknown parameter is the unit quaternion $x$ encoding a rotation of the source relative to the target. We consider an example from structural bioinformatics and test MCMC samplers to tackle this rigid registration task. Adenylate kinase (AK) undergoes a large conformational change upon ligand binding. This internal structural change cannot be modeled as a global rigid transformation. Only the core domain, comprising a 60\%{} subset of all points, can be related to each other via a rigid transformation. In our tests, the target and source point clouds show the closed and open conformation of AK (the protein data bank (PDB) codes 1AKE and 4AKE; see Fig. \ref{fig:proteins}). We fix $\sigma=1$\AA{} and $\omega=0.4$ (motivated by the fact that the core domain contains roughly 60\%{} of all points); the bounding box volume of 1AKE is $5.73 \times 10^{4}${\AA}$^3$. To guide the visual inspection, we use the {\em chainbow} coloring from blue to red offered by the Pymol software. Note that the posterior (\ref{eq:posterior-cpd}) is invariant under reordering of the points. 
  
\begin{figure}[bt]
  %\centering\includegraphics[width=\textwidth]{figures/protein_registration_iterations}
  %
  %\centering\includegraphics[width=\textwidth]{figures/protein_registration_times}
  %
  \centering\includegraphics[width=\textwidth]{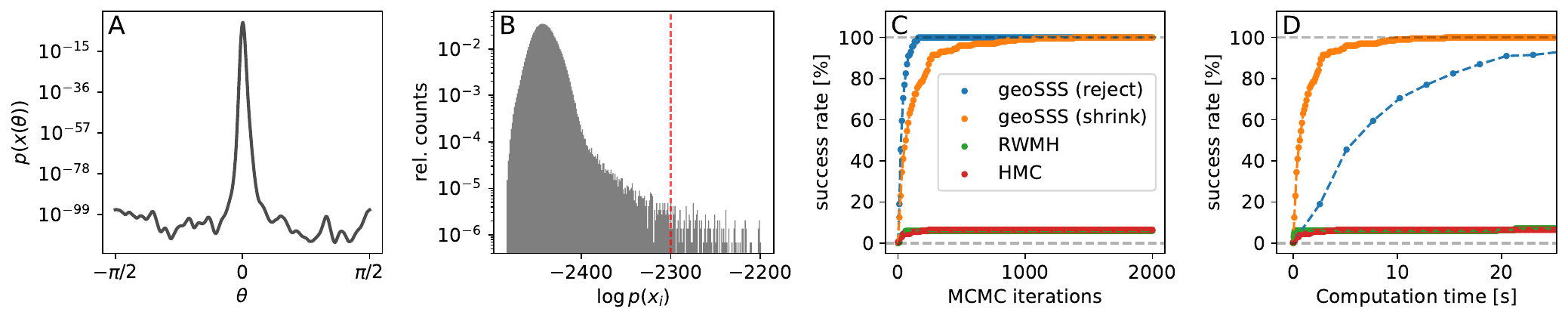}
  \caption{Probabilistic rigid registration of the open and closed structure of adenylate kinase based on posterior $p$ given in \eqref{eq:posterior-cpd}. (A) Slice through $p$ along a great circle $x(\theta) = \cos\theta\, x_1 + \sin\theta\, x_2$ where $x_1\in\sphere{3}$ is the global optimum and $x_2\in\subsphere{2}{x_1}$. (B) Marginal distribution of log posterior probabilities $\log p(x_i)$ based on a discretization of $\sphere{3}$ using $\sim 1.28 \times 10^6$ regularly placed unit quaternions $x_i \in \sphere{3}$. (C) Success rate against number of Markov chain iterations for all four samplers. (D) Success rate against computation time. %the geodesic slice sampler based on the shrinkage strategy and RWMH. Note that the ticklabels of the $x$ axis indicate the number of slice sampling iterations, the number of RWMH iterations was 5 times larger in order to match the computational effort of running both samplers.
  }\label{fig:cpd}
\end{figure}

The posterior density (\ref{eq:posterior-cpd}) has many local peaks as is clear from Figure \ref{fig:cpd}A showing a slice through $p$ along a random great circle passing through the posterior mode. Rigid registration by sampling from posterior (\ref{eq:posterior-cpd}) is challenging, because standard MCMC samplers tend to get stuck in subordinate peaks and miss the global maximum. To gain an overview of the posterior, we systematically evaluate $p$ over a regular tessellation of $\sphere{3}$ based on the 600-cell \citep{Straub17}. The 600-cell is the 4D analog of the regular icosahedron and can be used to approximate $\sphere{3}$ at varying degrees of resolution. A single cell is a 3D tetrahedron whose corners lie in $\sphere{3}$. By splitting each tetrahedron into eight equally-sized tetrahedra we obtain a tessellation at a higher resolution. Starting from the 600-cell encoding 300 unique rotations, we repeat the splitting process four times to obtain 1279264 unit quaternions that tessellate $\sphere{3}$ in a regular fashion. Evaluation of $p$ at these quaternions offers an accurate discretization of the posterior that allows us to study its shape (see Figure \ref{fig:cpd}B). The global maximum of $\log p$ on the quaternions of the 600-cell tessellation is $-2192.89$. Although, an overwhelming fraction of quaternions achieves only very small posterior probabilities, almost all probability mass is concentrated in the maximum posterior peak: the fractional volume of $\sphere{3}$ covered by quaternions $x\in\sphere{3}$ with $\log p(x) > -2300$ is close to zero ($1.27\times 10^{-4}$ based on the 600-cell tessellation), but the cumulative probability is close to one ($1 - 4 \times 10^{-44}$). In hindsight, due to the high concentration of the target distribution one could argue that the ability to identify the mode, effectively serving as an optimizer, is of primary importance, while posterior-based uncertainty quantification is less relevant.
% We compare the spherical slice sampler using a shrinkage strategy to RWMH.}

To investigate the robustness of probabilistic rigid registration with the geodesic slice samplers, RWMH and HMC, we launched 200 MCMC chains starting from random initial orientations that are generated by sampling unit quaternions uniformly from $\sphere{3}$. We assess the performance of the samplers by counting how many among all 200 runs reached a log posterior probability greater than $-2300$ and consider these simulations a ``success'', because it located the dominant posterior mode rather than getting stuck in a subordinate peak. As shown in Figure \ref{fig:cpd}C, the fraction of successful slice sampling runs increases with the number of MCMC iterations. As expected, the rejection strategy converges faster to the posterior mode than the shrinkage strategy, because it explores the sample space more exhaustively. The success rate of the other two MCMC samplers, RWMH and HMC, remains constantly low reaching only up to 7\%{} after 2000 iterations. This demonstrates that geodesic slice sampling can deal with multi-modal posteriors, whereas RWMH and HMC tend to get stuck independent of the number of MCMC iterations. Both geodesic slice samplers reach a success rate of greater than 50\%{} for about 50 iterations. The shrinkage strategy achieves 100\%{} success rate after 1500 iterations, whereas the rejection strategy is at 100\%{} success rate already after 200 iterations. In contrast, the performance of RWMH and HMC is quite poor with a success rates ranging between 3--7\%{}. A similarly poor performance is also observed for a custom-made Gibbs sampler (not shown).

Rigid registration is often just one of many sampling tasks within the context of a more complex probabilistic model. Therefore, it is desirable that highly probable rotations are sampled after few MCMC iterations within a short computation time. This is not achieved by the rejection strategy (see Fig. \ref{fig:cpd}D), because the dominant posterior mode is rather narrow resulting in many rejections. The shrinkage strategy, although requiring more steps to successfully detect the posterior peak with high probability, clearly outperforms any of the other samplers when taking also the computation time into consideration.

\subsection{Mixture of von Mises-Fisher Distributions}
To study the performance on a multi-modal target, we test the slice samplers as well as RWMH and HMC also on a $K$-component mixture model of von Mises-Fisher (vMF) distributions in $\dim$ dimensions. The vMF distribution is defined by the unnormalized density
\begin{equation}\label{eq:vMF}
	\vMF{}(x; \mu, \kappa) = \exp (\kappa\mu^T\!x),\quad x\in\sphere{\dim-1},
\end{equation}
where $\kappa>0$ is the concentration parameter and $\mu \in \sphere{\dim-1}$. For a given $K\in\mathbb{N}$ our target distribution is a  mixture of vMF distributions where each component has the same weight $1/K$ and a shared concentration parameter $\kappa$, i.e., the corresponding unnormalized density takes the form (note that the normalizers of each component are identical due to the shared concentration parameter)
\begin{equation}\label{eq:mix-vMF}
	p_{\text{mix}}(x) =  \sum_{k=1}^K \vMF(x; \mu_k, \kappa), \quad 
	x\in\sphere{\dim-1},
\end{equation}
where every $\mu_k \in \sphere{\dim-1}$ with $k\in\{1,\dots,K\}$ is sampled w.r.t. the uniform distribution on the unit sphere and then fixed. 
%To generate a baseline, we analytically compute the marginal distribution from the mixture of von Mises-Fisher distributions.
%
%\added{We also construct another baseline sampler suitable for such multimodal targets. A Markov kernel mixture combining a well-tuned RWMH for local exploration (within a given mode) and an independence sampler for global jumps (between modes) is introduced. This can be given as}
%%
%\begin{equation}\label{eq:kernel-mixture-sampler}
%	H_{\text{mix}}(x, \cdot) = \alpha H_{\text{RWMH}}(x, \cdot) + (1 - \alpha) H_{\text{Indep.}} (x, \cdot),
%\end{equation}
%%
%\added{where $H_{\text{mix}}$ is the Markov kernel mixture, $H_{\text{RWMH}}$ and $H_{\text{Indep.}}$ correspond to the spherical variants of the RWMH and independence kernel respectively. Here we denote $\alpha \in [0, 1]$ as a ``mixing probability" hyperparameter that determines which kernel is used for a given state. For implementation details refer to algorithm~\ref{alg:kernel-mixture-sampler} under appendix~\ref{app:rwmh}.} 

		For this target distribution we also consider a Metropolis algorithm with mixture proposal based on a  random walk and independent sampling, which is referred to as mixture-MH. The random walk part allows local exploration, whereas the independence sampling, regarding the uniform distribution on the sphere, encourages global moves. In this scheme the mixture hyperparameter $\alpha \in [0,1]$ appears: With probability $\alpha$ a random walk and with probability $1-\alpha$ an independent sampling proposal is used. The resulting 
		Metropolis kernel can also be interpreted as a mixture of RWMH and independent Metropolis. For implementation details we refer to Algorithm~\ref{alg:kernel-mixture-sampler} under Appendix~\ref{app:rwmh}.
		
\begin{figure}[bt]
	\centering{
		\includegraphics[width=\textwidth]{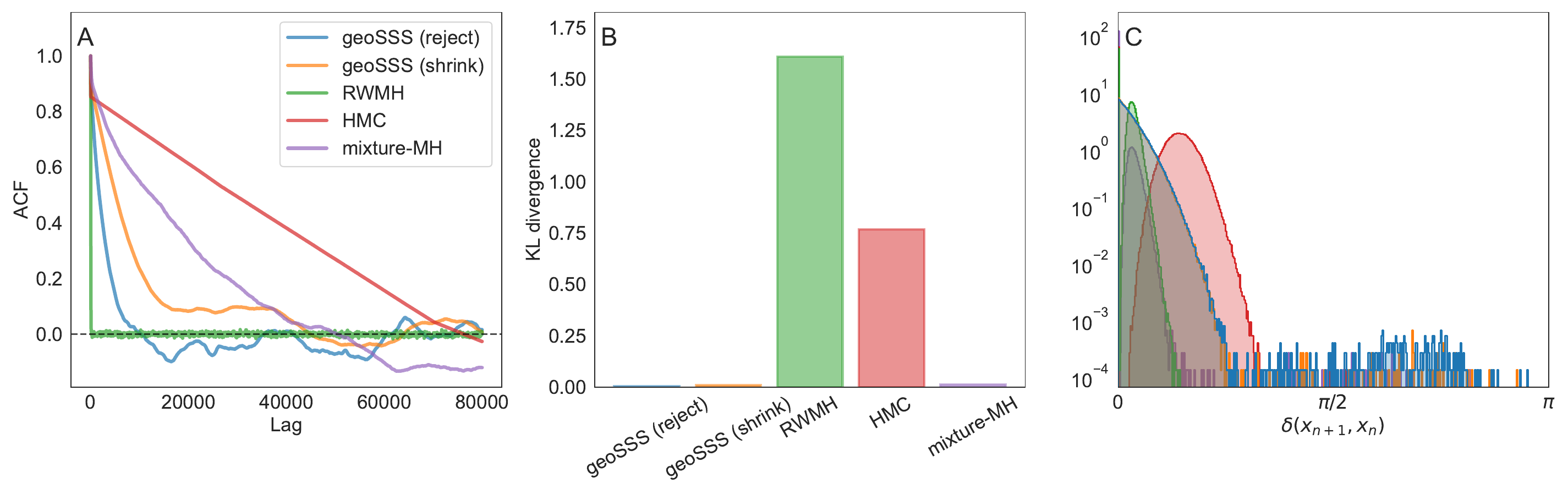}
	}
	\caption{(A) ACF plotted for the first dimension from the $10$-dimensional mixture of vMF distributions with $K=5$ and $\kappa=100$. (B) Kullback-Leibler divergence between the empirical and ideal frequency with which each mode is visited by the four MCMC samplers. (C) Geodesic distance (log-scale) between successive approximate samples.} \label{fig:mix-vMF-d10-acf-kld-dist}
\end{figure}

	We evaluate both geoSSS variants against RWMH,  HMC and the mixture-MH (for suitable $\alpha\in [0,1]$) on a $10$-dimensional mixture of vMF distributions with $K=5$ vMF-mixture components on the unit sphere $\sphere{9}$. We set $\kappa=100$ and draw $10^6$ samples. 
\begin{figure}[bt]
	\centering{
		\includegraphics[width=\textwidth, trim=0cm 30cm 0cm 0cm, clip]{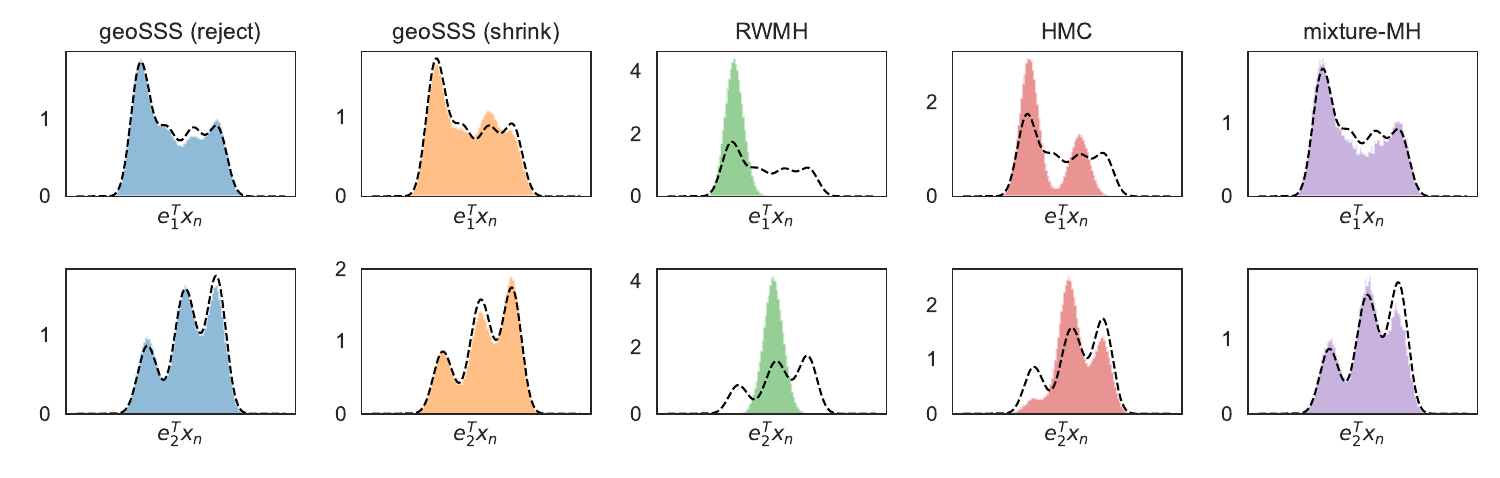}
	}
	\caption{Marginal distributions (shown in colors) for the mixture of vMF distributions ($\dim=10, \kappa=100, K=5$) where columns corresponds to MCMC methods and rows correspond to the dimension of the marginal (only the first two dimensions are shown). Here, $\{e_i\}_{i=1}^{10}$ denotes the standard basis in $\reals{10}$. The black dashed line indicates the true marginals computed analytically for the mixture of vMF distribution.} \label{fig:mixvMF-d10-hist}
\end{figure}
To gain some insight into the performance of the samplers on the multi-modal target, we first look at the marginal distributions (see Fig.~\ref{fig:mixvMF-d10-hist}). We estimate marginal histograms from the samples and compare them with the marginal distributions of the vMF mixture that we computed analytically. Both variants of geoSSS represent the underlying mixture significantly better compared to HMC and especially RWMH which seems to explore only a single component of the vMF mixture. The mixture-MH is almost on-par with our GeoSSS (shrink) variant. However, this represents the best result achieved by sweeping through $\alpha$ values\footnote{For the $\dim=10, \kappa=100, K=5$ vMF mixture target, the best mixing probability $\alpha=0.2$ corresponds to the lowest KL divergence value (see Fig.~\ref{fig:mix-vMF-d10-acf-kld-dist}B). For the systematic evaluation, we varied $\alpha \in [0,1]$ with a grid spacing of $0.1$.} with $10^6$ MCMC steps for each case and therefore represents an additional computational burden of tuning this hyperparameter along with the step-size of RWMH for every target density.

Next we study quality measures to obtain more quantitative insights into the sampling performance. RWMH is the fastest to decorrelate, which is reflected by the rapid decay of the autocorrelation function (ACF) (see Fig.~\ref{fig:mix-vMF-d10-acf-kld-dist}A). However, the reason for the fast decorrelation is that RWMH fails to explore the entire distribution, but is stuck in a single component. The geoSSS variants decorrelate much faster as compared to mixture-MH and HMC. Note that the slow, linear decay of the ACF for HMC results from the fact that it spends a large fraction of the simulation time in a single mode and only occasionally escapes to another mode (also HMC only visits three out of all five modes in the course of the simulation).

As a measure for the quality of approximate sampling we consider the frequency with which the samplers visit the $K$ modes of the target mixture model (\ref{eq:mix-vMF}). Ideally, these frequencies should be constant, because each component of the mixture has the same weight and concentration parameter. We contrast the empirical frequency $q_k$ with which the $k$-th mode is visited by a sampler with the uniform distribution by using the Kullback-Leibler (KL) divergence
\begin{align*}
	\text{KL}(q\mid \ell) := \sum_{k=1}^K q_k \log (q_k / \ell_k)
\end{align*}
where $\ell = (\ell_1, \ldots, \ell_K)$ with $\ell_k= 1/K$ for $k \in \{1, \ldots, K\}$ is the ideal distribution and $q = (q_1, \ldots, q_K) \in [0,1]^K$. To estimate $q_k$, we assign each sample $x_n$ to the component under which $x_n$ is most likely, i.e.,
\[
	k_n=\underset{k\in \{1, \ldots, K\}}{\text{argmax}}\, \vMF(x_n; \mu_{k}, \kappa).
\]
Based on the allocations $k_n$, we can count how often each mode is visited by the Markov chain. A good MCMC sampler should produce $q_k \approx 1 / K$ resulting in $\text{KL}(q\mid \ell) \approx 0$. The more the KL divergence differs from the minimum value of zero, the greater is the mismatch between the ideal and empirical frequency of mode visits. 
As seen from  Fig.~\ref{fig:mix-vMF-d10-acf-kld-dist}B, geoSSS (reject) achieves the lowest KL divergence whereas geoSSS (shrink) and mixture-MH perform comparably in achieving low KL divergence values. However, RWMH and HMC produce large KL divergences, indicating a complete failure to detect some of the components and, as a result, a poorer representation of the target distribution compared to geoSSS. Indeed, RWMH is stuck in a single mode in the course of the simulation. HMC fares better, but also misses two out of five modes.

Another quality measure is the geodesic or great circle distance between successive samples given as
\[
\delta(x_{n+1}, x_n) := \arccos(x_{n+1}^T x_n)\, .
\]
An efficient MCMC algorithm should explore the sphere rapidly by making large leaps from one sample to the next.
The great circle distance between successive samples (see Fig. \ref{fig:mix-vMF-d10-acf-kld-dist}C) shows that the slice samplers and mixture-MH explore the sphere more efficiently than HMC and RWMH.

\begin{figure}[tb]
	\centering{
		\includegraphics[width=\textwidth]{
			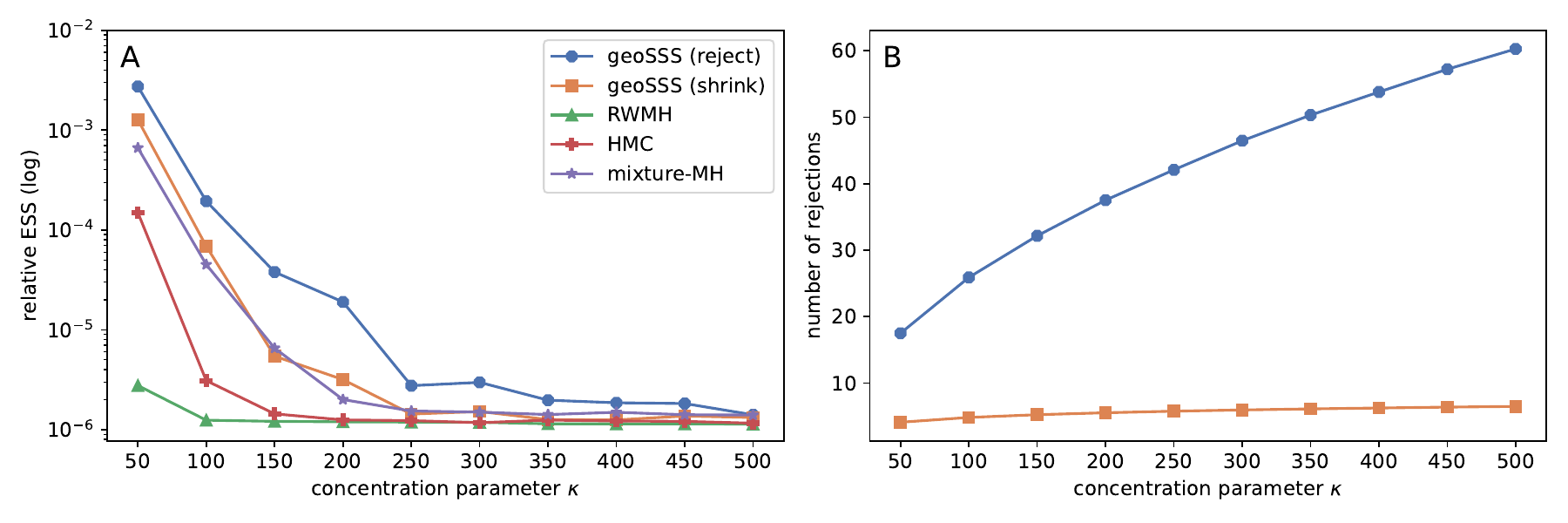}
	}
	\caption{Comparing the MCMC samplers for the mixture of vMF distribution ($d=10$, $K=5$) by varying $\kappa$ from 50 to 500. (A) Considering the first dimension, estimating the relative effective sample size (ESS) for 10 chains. (B) Number of rejections per MCMC step. } \label{fig:ess-rejections-mixvMF-d10-kappa50_500}
\end{figure}

We further evaluate the performance of the samplers by fixing the components on the sphere and increasing the concentration parameter $\kappa$ from $50$ to $500$, therefore making the distributions ``spikier''. We estimate the effective sample size (ESS) for each method by considering $10$ chains and $10^6$ MCMC steps from the first dimension, resulting in ESS values per method for each $\kappa$. Figure \ref{fig:ess-rejections-mixvMF-d10-kappa50_500}A shows the estimated relative\footnote{Here and elsewhere the relative ESS refers to the estimated ESS divided by the total number of MCMC steps.} ESS  for each of the four samplers as a function of the concentration parameter $\kappa$. Overall, we observe that sampling the mixture model becomes more challenging as the components become more concentrated with increasing $\kappa$ for all methods. Notably, the geoSSS (reject) variant achieves the best relative ESS. We can again observe the trend that geoSSS (shrink) and mixture-MH (with the mixture hyperparameter fine-tuned for every $\kappa$ corresponding to the highest relative ESS) are comparable as well as that both  outperform RWMH and HMC, particularly for lower $\kappa$ values. At large $\kappa$ values, all samplers perform similarly poorly and fail to produce a reliable approximation of the target.
It is worth noting that although the rejection-based geoSSS yields the highest ESS, it is at the cost of generating a significantly larger number of rejected samples in comparison to the shrinkage-based geoSSS. This is demonstrated in Fig.~\ref{fig:ess-rejections-mixvMF-d10-kappa50_500}B, where it is observed that the rejection-based geoSSS generates an increasingly larger number of rejected samples for higher values of $\kappa$. For instance, for $\kappa=50$, approximately 17 rejections happen per MCMC step, while for $\kappa=500$, around 60 rejections occur. On the other hand, the number of rejections for the shrinkage-based geoSSS increases at a much slower rate with $\kappa$. It can be seen here that for $\kappa=50$, we have approximately 4 rejections per MCMC step, and for $\kappa=500$, approximately just 6 rejections per MCMC step occur. 
The number of rejections directly determines the number of log probability evaluations and thereby the computational costs of the slice samplers. 
RWMH and HMC, on the other hand, have a fixed computational budget per MCMC step. 
In case of RWMH, one step requires a single log probability evaluation, whereas an HMC step involves one log probability evaluation as well as multiple gradient evaluations, one for each leapfrog step. In our settings, we use 10 leapfrog integration steps. 
In summary, RWMH has the lowest computational costs, but also the smallest ESS. The computational costs of shrinkage-based geoSSS is smaller than the costs of HMC, whilst it also performs better than HMC in terms of the other evaluation criteria.
\\[-4ex]
\subsection{Curved Distribution on the Sphere}
For a target density where the `mass' accumulates along narrow connected regions, HMC is expected to show superior performance. To study such a sampling problem, we define a spherical distribution along a curve. 
The curve is created by picking 10 consecutive points on the $(d-1)$-sphere and to ensure a smooth path, we connect every two successive points $x, y \in \sphere{\dim -1}$ using spherical linear interpolation (slerp), i.e., we apply  
%\shantanu{Making this statement more general because we don't use TSP for $d>3$ curves and the curve generation method is also different in that case.}
\[
%	\added{\colon[0,1] \to \sphere{d-1},} \qquad 
	t \mapsto \frac{\sin(\theta\, (1-t))x + \sin(\theta\, t) y}{\sin(\theta)}, 
\]
as map from $[0,1]$ to $\sphere{d-1}$, where $\theta=\arccos(x^Ty)$, cf. \citep{Hanson95}. If $\mu(t)$ with $t \in [0, 1]$ 
is the curve obtained by concatenating and rescaling the slerps between the 9 pairs of successive points, then we define the spherical distribution with unnormalized density
\begin{equation}\label{eq:curved-vMF}
	p(x) = \exp\left(\kappa \max_{t\in[0, 1]} x^T\!\mu(t) \right),
	\qquad x\in  \sphere{\dim -1}, 
\end{equation}
which we also call curved von Mises-Fisher distribution (curved vMF). 
The maximum of $x^T\!\mu(t)$ for a single slerp can be computed in closed form. The total maximum over the entire curve is then the maximum over the maxima of slerps connecting two successive points. The probability distribution
corresponding to
\eqref{eq:curved-vMF} concentrates probability mass around the curve $\mu(t)$. As in the case for the standard vMF distribution, the parameter $\kappa > 0$ controls the concentration. 

For our initial sampling tests for the curved vMF target, we consider $\sphere{2}$ and  choose $\kappa=500$. 
Again, we run all four MCMC samplers. Rejection-based geoSSS and HMC\footnote{More details on the implementation of the required gradient can be found in the Python library \url{https://github.com/microscopic-image-analysis/geosss} and in the following note \href{https://github.com/microscopic-image-analysis/geosss/blob/main/scripts/gradient-curve.pdf}{\texttt{geosss/scripts/gradient-curve.pdf}}}  take roughly the same amount of time, with HMC taking slightly longer. However, both are almost twice as slow as shrinkage-based geoSSS. RWMH is by far the fastest method (refer to Appendix~\ref{supp-run-times} for run times). 
\begin{figure}[tb]
	\centering{\includegraphics[width=1\textwidth, trim=2.25cm 2cm 2.1cm 0cm, clip]{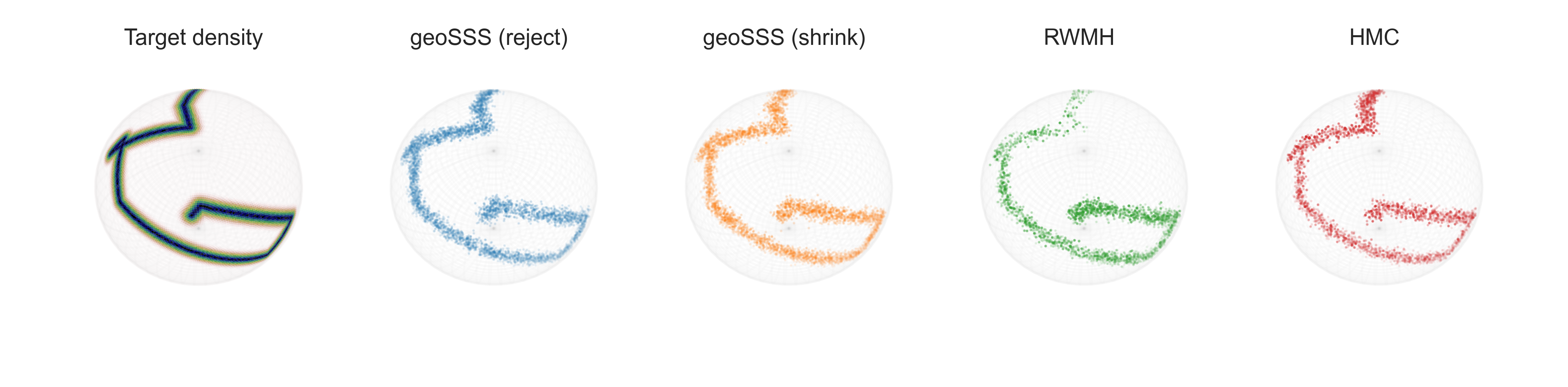}}
	\caption{First panel from the left shows the curved vMF target density on the $2$-sphere with concentration parameter $\kappa =500$. Remaining panels show the last 5000 samples from the four different MCMC samplers targeting this density.}\label{fig:curve-samples}
\end{figure}

\begin{figure}[tb]
	\centering{
		\includegraphics[width=\textwidth]{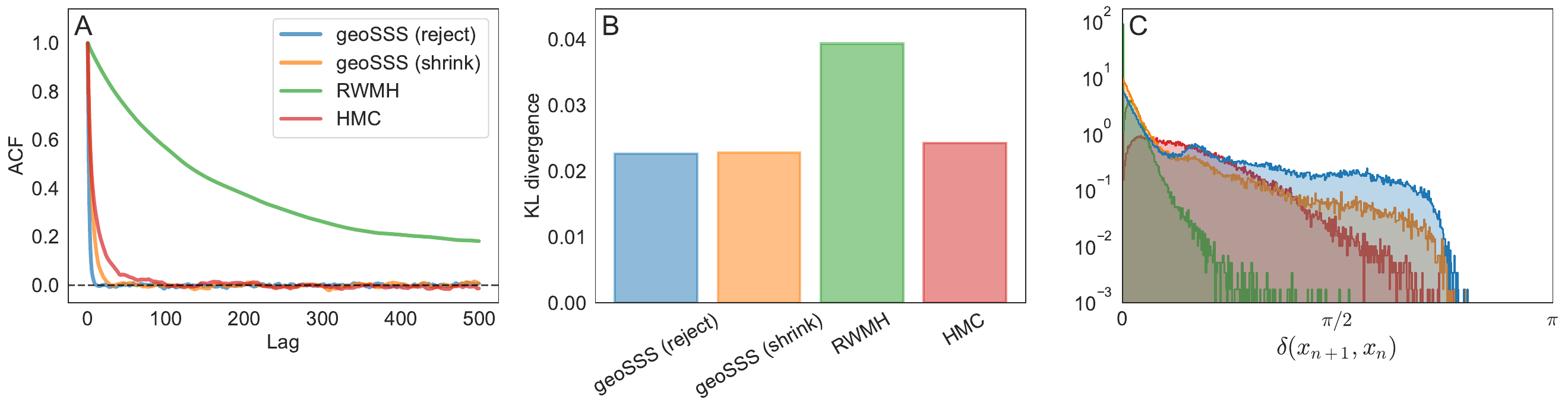}
	}
	\caption{(A) ACF of the first dimension plotted for the curved vMF target on $2$-sphere  based on $10^5$ samples. (B) KL divergence between a discretized version of the curved vMF target on 2-sphere and a histogram compiled from the MCMC samples. (C) Geodesic distance (log-scale) between successive approximate samples for the curved von Mises-Fisher distribution \eqref{eq:curved-vMF} on $2$-sphere.} \label{fig:curve-acf-kld-dist}
\end{figure}

Figure \ref{fig:curve-samples} shows the target density and samples of the four MCMC algorithms generated in the last 5000 iterations. The underlying target density shown here is evaluated by uniformly distributing $3 \times 10^4$ points on the $2$-sphere using the method of \cite{Saff1997}. Visual inspection shows that the slice samplers and also HMC explore the target much more rapidly than RWMH, which is also reflected in the high autocorrelation (see Fig. \ref{fig:curve-acf-kld-dist}A). To verify this observation, we  measure the discrepancy between the discretized target and the histogram generated by the sampling procedure using KL divergence (see Fig. \ref{fig:curve-acf-kld-dist}B). Moreover, we monitor the distance between successive samples (see Fig. \ref{fig:curve-acf-kld-dist}C). 
Across all three evaluation criteria  we consider, RWMH is clearly disfavored. On the other hand, geoSSS and HMC show a similar performance on this low-dimensional target.

\begin{figure}[tb]
	\centering
	\includegraphics[width=\textwidth]{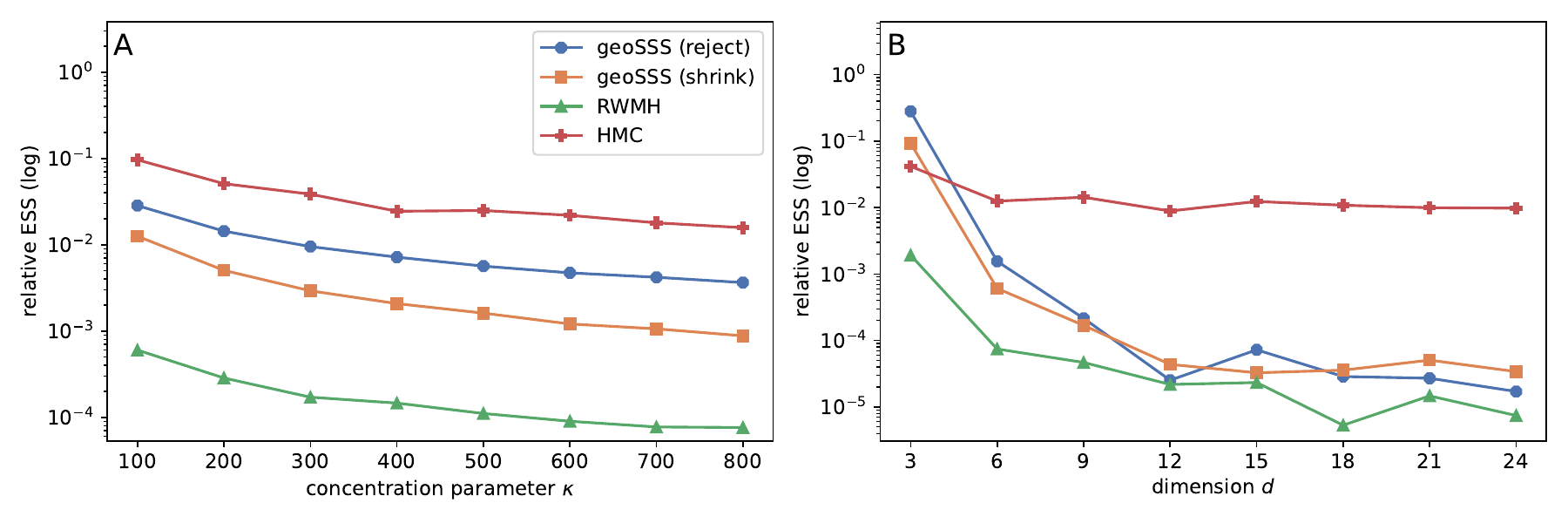}
	\caption{Comparing the relative ESS values. (A)  Fixing the dimension $d=5$ and varying the concentration paramater. (B) Fixing $\kappa=500$ and varying the dimensions.}
	\label{fig:ess_curve}
\end{figure}

  However, this picture changes as the dimensionality of the target increases from $d=3$ to $d=5$. 
  %We further study sampler performance, especially as the dimensionality of the curved vMF target increases. By drawing $10^6$ samples and fixing the dimension $d=5$ ($4$-sphere) with the concentration parameter $\kappa=800$, we create a target that is both higher dimensional and more narrowly concentrated. %For visualization, we use the scatterplot matrix from the \textit{corner} library (see \cite{corner}). As shown in Fig. \ref{fig:corner-plot}, this plot displays one- and two-dimensional projection (as $1$D and $2$D histograms) of the multidimensional samples, with $2$D histogram also showing scattered samples against the underlying target. From the $1$D histograms (shown along the diagonal), it is evident that all samplers, except RWMH, match closely.
  %  Examining the ACF for the first dimension (see left panel of Fig.~\ref{fig:curve-acf-dist-d5}), we observe that HMC decorrelates fastest, followed by geoSSS (reject), geoSSS(shrink), and, finally, RWMH, which decorrelates the slowest. The geodesic distance between successive samples (see right panel of Fig.~\ref{fig:curve-acf-dist-d5}) reveals that the slice samplers traverse larger distances compared to the other samplers, with RWMH only reaching a maximum distance of approximately $\pi / 8$. The maximum geodesic distance covered by HMC is still smaller than that of the slice samplers, suggesting that HMC does not require larger steps to fully explore the target distribution. To further confirm that HMC is the best sampler for such targets, we
  We compute the relative ESS for the first dimension by varying the concentration parameter. By using $10$ chains, fixing the dimension $d=5$ and drawing $10^6$ samples per chain, we observe (see Fig. \ref{fig:ess_curve}A), that HMC consistently outperforms the remaining samplers across all values of $\kappa$. The trend of decreasing ESS with increasing $\kappa$ aligns with results from the previous section on the mixture of vMF distributions.
%
%\begin{figure}
%	\centering{
%		%\includegraphics[width=0.9\textwidth]{figures/curve_corner_5d_kappa800.pdf}
%	}
%	\caption{\added{Scatterplot matrix representation for the curved vMF target for $4$-sphere and concentration parameter $\kappa=800$.}} \label{fig:corner-plot}
%\end{figure}
%
HMC is expected to perform exceptionally well on such targets, even if the dimensionality is increased further. To assess the performance of the remaining samplers under similar conditions, we repeat the ESS calculations, however in this case, we fix the concentration parameter with $\kappa=500$ and vary the dimensions from $3$ to $24$. As observed in Fig. \ref{fig:ess_curve}B, while the ESS values of the slice samplers are better than RWMH, all three exhibit rapid decline with increasing dimension. As expected, in contrast, HMC demonstrates superior performance, maintaining fairly consistent ESS values across all  dimensions.
%
%\begin{figure}
%	\centering{
%		\includegraphics[width=0.9\textwidth]{figures/curve_acf_dist_5d_kappa800.pdf}
%	}
%	\caption{\added{Left panel: ACF plotted for the curved vMF target on $4$-sphere with $\kappa=800$ and based on $10^6$ samples. Right panel: Geodesic distance (log scale) between successive approximate samples for the same target.}} \label{fig:curve-acf-dist-d5}
%\end{figure}
%
\section{Summary}
We introduce two slice sampling based MCMC-methods on the sphere, the ideal geodesic and geodesic shrinkage slice samplers, that use movements on great circles. For the Markov kernels of both samplers we are able to establish reversibility and explicit convergence statements under mild assumptions. 
In numerical experiments, we see that in particular for moderately concentrated target distributions with a moderate number of dimensions our slice samplers perform well and can compete with RWMH and HMC on the sphere, while having the additional advantage of being free of tuning parameters. 
For multimodal target distributions we observe that the ideal geodesic slice sampler and the geodesic shrinkage slice sampler even outperform RWMH and HMC. However, we observe that the performance of our methods deteriorates for increasing concentration and increasing dimension of the target distribution.  This dependence on concentration and dimension already appears in Theorem \ref{Thm: Uniform ergodicity ISS} and Theorem \ref{Thm: Uniform geometric ergodicity of SSS}. Regarding the comparison between the two geodesic slice samplers, the ideal version seems to outperform the shrinkage based sampler in terms of Markov chain transitions. However, this comes with a (significantly) greater computational cost per transition.
\vspace*{-2ex}                                                          

% Acknowledgements and Disclosure of Funding should go at the end, before appendices and references
\acks{\\[-5ex]
	All authors are grateful for the support of the DFG within project 432680300 -- SFB 1456 subprojects A05 and B02. We also thank Philip Sch\"ar for valuable comments and discussions about the topic. M. Habeck and S. Kodgirwar gratefully acknowledge funding by the Carl Zeiss Stiftung within the program ``CZS Stiftungsprofessuren'' and by the German Research Foundation (DFG) within grant HA 5918/4-1.
	M. Hasenpflug and D. Rudolf were supported by the German Research Foundation (DFG grant number 522337282).}

% This is not part of the standard template, but already published JMLR versions have done this, not capitalizing the first letter here as well.
\vspace*{-1ex}
\subsection*{Code Availability}
\vspace*{-0.5ex}
We have implemented our geoSSS algorithms, including the spherical RWMH and HMC methods (detailed in Appendices \ref{app:rwmh} and \ref{app:hmc}), in our {\ttfamily geosss} package available at \url{https://github.com/microscopic-image-analysis/geosss}. You can find the scripts used to generate the numerical illustrations in the \href{https://github.com/microscopic-image-analysis/geosss/tree/main/scripts}{\ttfamily scripts} directory, and the protein example data in the \href{https://github.com/microscopic-image-analysis/geosss/tree/main/data}{\ttfamily data} directory. Also the precomputed results are archived on Science Data Bank \url{https://doi.org/10.57760/sciencedb.30181}.


\clearpage

\appendix
\renewcommand\thefigure{A\arabic{figure}}
\renewcommand{\thetable}{A\arabic{table}}
\setcounter{figure}{0}
\setcounter{table}{0}

\section{Simulation of $\mathcal{U}_{\subsphere{\dim - 2}{x}}$}
\label{Sec: Sampling mu_x}
Implementing the slice sampling algorithms 
proposed
in this paper involves sampling from $\mathcal{U}_{\subsphere{\dim - 2}{x}}=\frac{1}{\area{\dim -2}}\,\unitangsphere{x}$ for $x \in \sphere{\dim -1}$. A scheme for performing this is provided in Algorithm \ref{alg:sample-subsphere} and justified as follows.
It exploits that $\subsphere{\dim - 2}{x}$ can be interpreted ``as tilted $\sphere{\dim -2}$, such that it lies in the $(\dim-1)$-dimensional hyperplane orthogonal to $x$''. We formalize the arguments and require further objects.
For $x \in \sphere{\dim -1}$ let 
\[
	E_x \coloneqq \{ y \in \reals{\dim} \mid \inner{y}{x} = 0\}
\]
be the $(\dim -1)$-dimensional hyperplane with normal vector $x$.
Since this is a $(\dim -1)$-dimensional linear subspace of $\reals{\dim}$, 
we can choose an orthonormal basis $e_1^{(x)}, \ldots, e_{\dim-1}^{(x)} \in E_x$ of $E_x$.
The isometric mapping 
\begin{equation}\label{Eq: Definition of alpha}
	\varphi_x: \reals{\dim-1}  \to E_x, \qquad (y_1, \ldots, y_{\dim-1}) \mapsto \sum_{i = 1}^{\dim-1} y_i e_i^{(x)}
\end{equation}
describes the rotation of $\reals{\dim -1}$ into $E_x$.
Therefore it maps the unit sphere in $\reals{\dim -1}$ to the unit sphere in $E_x$, i.e., $\varphi_x(\sphere{\dim -2}) = \subsphere{\dim -2}{x}$.
Isometries preserve volume, such that
\begin{equation}\label{Eq: mu_x as image measure}
	\unitangsphere{x}(A) = \unisphere{\dim -2}\left( \varphi_x^{-1}(A) \right), \qquad \forall \, A \in \sigmaalgebra{\subsphere{\dim  -2}{x}},
\end{equation}
provides the crucial relation
between the volume measure on $\sphere{\dim -2}$ and the volume measure $\unitangsphere{x}$ on the tilted version $\subsphere{\dim -2}{x}$,
\citep[see also][Exercise 25.4]{Munkres}. 
In particular, $\unitangsphere{x}(\subsphere{\dim -2}{x}) = \area{\dim -2}$.

We prove the validity of Algorithm~\ref{alg:sample-subsphere} based upon this relationship between $\unisphere{\dim -2}$ and $\unitangsphere{x}$.
For $x \in \sphere{\dim -1}$ we
establish an expression for the mapping $y \mapsto y - (x^T\!y)\, x$ in terms of $\varphi_x$ that describes the projection of $\mathbb{R}^\dim$
onto $E_x$. Then, we deduce that Algorithm~\ref{alg:sample-subsphere} realizes a transformed normalized standard normally distributed random variable under the isometry $\varphi_x$. By \eqref{Eq: mu_x as image measure} this implies the desired result formalized as follows.
\begin{algorithm}[b]
	\caption{Sampling from $\mathcal{U}_{\subsphere{\dim - 2}{x}}$.}
	\label{alg:sample-subsphere}
	\begin{algorithmic}[1]
		\INPUT point $x \in \sphere{\dim-1}$
		\OUTPUT sample $v$ from $\mathcal{U}_{\subsphere{\dim - 2}{x}}$
		\STATE Draw $Y \sim \normal{0}{\mathrm{Id}_\dim}$, call the result $y$.
		\STATE Set $z = y - (\inner{x}{y})\, x$ and $v = z / \|z\|$.
	\end{algorithmic}
\end{algorithm}
\begin{lemma}
	Let $Y$ be a random variable distributed according to the $\dim$-dimensional standard normal distribution $\normal{0}{\mathrm{Id}_\dim}$. Then, for any $x \in \sphere{\dim -1}$ we have $\frac{Y - \inner{x}{\,Y}\, x}{\norm{Y - \inner{x}{\,Y}\, x}}\sim
	\mathcal{U}_{\subsphere{\dim - 2}{x}}$.
\end{lemma}
\begin{proof}
	Let $U \in  \mathbb{R}^{\dim \times (\dim -1)}$ be the matrix with columns  $e_1^{(x)}, \ldots, e_{\dim -1}^{(x)}$ and let $\widetilde{U} \in \mathbb{R}^{\dim \times d}$ be the matrix with columns $e_1^{(x)}, \ldots, e_{\dim -1}^{(x)}, x$, where  $e_1^{(x)}, \ldots, e_{\dim -1}^{(x)}$ are defined as above. 
	Since $e_1^{(x)}, \ldots, e_{\dim -1}^{(x)}, x$ is an orthonormal basis of $\mathbb{R}^\dim$, we may write $y \in \mathbb{R}^\dim$ as
	$
	y = \sum_{i = 1}^{\dim -1} \alpha_i e_i^{(x)} + \alpha_{\dim} x
	$
	where $\alpha \coloneqq (\alpha_1,\dots,\alpha_d)^T = \widetilde{U}^Ty$. 
	Therefore, we have
	\begin{align*}
		y - (\inner{x}{y})\, x & = \sum_{i = 1}^{\dim -1} \alpha_i e_i^{(x)} 
		= \varphi_x \left((\alpha_1,\dots,\alpha_{d-1})^T\right) = \varphi_x (U^T y).
	\end{align*}
	Since $Y \sim \normal{0}{\mathrm{Id}_\dim}$ and $\widetilde{U}$ is orthogonal, also $\widetilde{U}^T Y \sim \normal{0}{\mathrm{Id}_\dim}$. Observe that the distribution of $U^TY$ is a marginal distribution of the distribution of $\widetilde{U}^T Y$. Hence $U^T Y \sim \normal{0}{\mathrm{Id}_{\dim-1}}$. This implies $\frac{U^T Y}{\norm{U^T Y}} \sim \mathcal{U}_{\sphere{\dim-2}} = \frac{1}{\area{\dim -2}}\, \unisphere{\dim -2}$. Using \eqref{Eq: mu_x as image measure} and  that $\varphi_x$ is a linear isometry, we obtain for $A \in \sigmaalgebra{\subsphere{\dim -2}{x}}$ that
	\begin{align*}
		\mathbb{P}\left( \frac{Y - \inner{x}{Y}\, x}{\norm{Y - \inner{x}{Y}\, x}} \in A \right)
		& = \mathbb{P}\left( \varphi_x \left(\frac{U^T Y}{\norm{U^T Y}}\right) \in A \right)
		=  \mathbb{P}\left( \frac{U^T Y}{\norm{U^T Y}} \in \varphi_x^{-1} (A) \right)         \\
		& = \frac{1}{\area{\dim -2}}\, \unisphere{\dim -2}\left( \varphi_x^{-1} (A) \right)
		= \frac{1}{\area{\dim -2}}\, \unitangsphere{x}(A) = \mathcal{U}_{\subsphere{\dim - 2}{x}}(A). 
		\\[-10ex]
	\end{align*}
\end{proof}

\section{Proof of Lemma~\ref{L: T indentities geodesic} and Lemma~\ref{L: Liouville measure invariant under T}}\label{Sec: Liouville measure invariant under T}
We start with the proof of Lemma~\ref{L: T indentities geodesic} that follows by an elementary calculation.
\vspace*{0.5ex}
\begin{proof}\emph{(Proof of Lemma \ref{L: T indentities geodesic}.)}
	For $x \in \sphere{\dim -1}$, $v \in \subsphere{\dim -2}{x}$ and $\theta \in \reals{}$,
	using classic trigonometric identities of sine and cosine, we obtain for all $r \in \mathbb{R}$ that
	\begin{align*}
	\geodesic{T_\theta(x,v)}(r) & = \cos(r) \big( \cos(\theta)x + \sin(\theta) v \big) + \sin(r) \big(\sin(\theta) x - \cos(\theta) v \big) \\
	& = \cos(\theta-r) x + \sin(\theta-r) v
	= \geodesic{(x,v)}(\theta - r).
	\end{align*}
	This implies $\geodesic{T_\theta(x,v)}(\theta) = \geodesic{(x,v)}(0) = \cos(0) x + \sin(0) v = x$.
\end{proof}	
We turn to the proof of Lemma~\ref{L: Liouville measure invariant under T}.	
For convenience we introduce
\begin{equation}\label{D: Definition of the Liouville measure}
	\liouville(A) := \int_{\sphere{\dim-1}} \int_{\subsphere{\dim-2}{x}} \mathbbm{1}_A(x,v) \unitangsphere{x}(\d v) \unisphere{\dim -1}(\d x),
	\qquad A \in \sigmaalgebra{\tangentbundle},
\end{equation} 
for the measure on $\tangentbundle = \bigcup_{x \in \sphere{\dim-1}} \big(\{x\} \times \subsphere{\dim-2}{x}\big)$ that ``sews'' up the volume measure on the fibers of $\tangentbundle$ by the volume measures on $\sphere{\dim -1}$, and call $\liouville$ the \emph{Liouville measure}.
We also use the following map.
\begin{definition}\label{D: Definition of the geodesic flow}
	Let $\theta \in \mathbb{R}$. The function
	\begin{align*}
		\gflow{\theta}: \tangentbundle  \to\tangentbundle, \qquad                                                             
		(x,v)                           \mapsto \big( \cos(\theta)x + \sin(\theta)v , -\sin(\theta) x + \cos(\theta) v\big)
	\end{align*}
	is called the geodesic flow on the sphere.
\end{definition}
To prove Lemma \ref{L: Liouville measure invariant under T} we exploit that we can write $T_\theta$ as a composition of the geodesic flow and a ``sign flip'' in the second component.
Then Lemma \ref{L: Liouville measure invariant under T} follows by invariance properties of the Liouville measure.
Note that when naming $\liouville$ and $\gflow{\theta}$ we adhere to the terminology of Riemannian geometry.
\vspace{0.5ex}
\begin{proof}\emph{(Proof of Lemma \ref{L: Liouville measure invariant under T}.)}
	Let $\theta \in \mathbb{R}$, and note that the Liouville measure is invariant w.r.t. the geodesic flow, that is,
	\[
	\liouville\left( \gflow{\theta}^{-1}(A) \right) = \liouville(A), \quad \forall A \in \sigmaalgebra{\tangentbundle},
	\]
	\citep[see e.g.,][Section V.2]{Chavel}.
	Moreover, define
	\begin{align*}
		\iota : \tangentbundle & \to \tangentbundle,\qquad
		(x,v) \mapsto (x, -v).
	\end{align*}
	Observe that the Liouville measure is invariant under $\iota$ \citep[see e.g.,][Lemma 1.34]{Paternain}. 
	We can express the map $T_\theta$ as $T_\theta = \iota \circ \gflow{\theta}$. 
	Therefore the invariance of the Liouville measure under $\gflow{\theta}$ and $\iota$ yields
	\begin{align*}
		\int_{ \tangentbundle} F\left( T_\theta(y) \right)\, \liouville(\d y)
		&=  \int_{\tangentbundle}(F\circ \iota \circ \gflow{\theta}) (y) \, \liouville(\d y) 
		= \int_{\tangentbundle} (F\circ \iota) (y) \, \liouville(\d y)\\
		&= \int_{\tangentbundle} F (y) \, \liouville(\d y).\\[-9ex]
	\end{align*}
\end{proof}

\section{Proof of Proposition~\ref{prop: rev_iss}}\label{sec: rev_iss}

A useful tool for showing reversibility of Markov kernels exhibiting the same structure as the ideal geodesic slice sampling kernel is Lemma 1 by \citet{HybridSliceSampling}. It applies to subsets of $\mathbb{R}^\dim$ in its original formulation, but can be extended to arbitrary $\sigma$-finite measure spaces. For the convenience of the reader we adapt the relevant parts of the aforementioned lemma to our setting.

\begin{lemma}\label{L: What to show for reversibility}
	Let
	\[
	P(x,A) \coloneqq \frac{1}{\targetdensity(x)} \int_0^{\targetdensity(x)} P_t(x,A\cap L(t))\, \d t, \qquad x \in \sphere{\dim -1}, A \in \sigmaalgebra{\sphere{\dim -1}}	,
	\]
%	be a Markov kernel where 
%	$P_t: \sphere{\dim -1}\times \sigmaalgebra{\sphere{\dim -1}} \to [0, 1]$ for $t \in (0, \norm{\targetdensity}_\infty)$ are themselves Markov kernels.
be a Markov kernel where 
$P_t: L(t)\times \sigmaalgebra{L(t)} \to [0, 1]$ for $t \in (0, \norm{\targetdensity}_\infty)$ are themselves Markov kernels.  
	If $P_t$ is reversible with respect to 
%		$\frac{1}{\unisphere{\dim -1}(\levelset{t})}\,\unisphere{\dim -1}\vert_{\levelset{t}}$ 
	$\mathcal{U}_{L(t)}$
	for all $t \in (0, \norm{\targetdensity}_\infty)$, 
	then $P$ is reversible with respect to $\target$.
\end{lemma}
\begin{remark}
 By the fact that $p$ is lower semicontinuous 
 for any $t\in (0,\norm{\targetdensity}_\infty)$ we have that $L(t)$ is open and non-empty. Therefore, $\unisphere{\dim-1}(L(t))\in (0,\infty)$ such that $\mathcal{U}_{L(t)}$ is well defined.
\end{remark}
We add an auxiliary result w.r.t. the volume of the geodesic level sets under the map $T_\theta$.
\begin{lemma}\label{L: T indentities - tau of level set under T}
		Let $x \in \sphere{\dim-1}$, $v \in \subsphere{\dim -2}{x}$ and $t \in (0, \norm{\targetdensity}_\infty)$. Then,
		for all $\theta \in \mathbb{R}$ we have ${\leb}\big(\glevelset{T_\theta(x}{v)}{t} \big) = {\leb}\big(\glevelset{x}{v}{t}\big)$.
\end{lemma}
	\begin{proof}
	Let $x \in \sphere{\dim -1}$, $v \in \subsphere{\dim -2}{x}$ and $t \in (0, \norm{p}_\infty)$.
	For $\theta \in \mathbb{R}$ set
	\begin{align*}
	w_\theta \colon [0, 2 \uppi) & \to [0,2\uppi),\quad
	r \mapsto (\theta-r) \cdot 
%	\mathbbm{1}_{\theta \geq r} 	
	\mathbbm{1}_{(-\infty,\theta\,]}(r)
	+  (\theta-r + 2 \uppi) \cdot 
%	\mathbbm{1}_{\theta < r}
	\mathbbm{1}_{(\theta,\infty)}(r)
	.
	\end{align*}
	Due to the $2 \uppi$-periodicity of sine and cosine, we have $\glevelset{T_\theta(x}{v)}{t} = w_\theta^{-1}\big(\glevelset{x}{v}{t}\big).$
 We obtain
	$
	{\leb}\left( \glevelset{T_\theta(x}{v)}{t} \right) = {\leb} \left(  w_\theta^{-1}\big(\glevelset{x}{v}{t}\big) \right)
	= {\leb} \left(  \glevelset{x}{v}{t} \right),
%		\\[-6ex]
	$
		since the Lebesgue measure ${\leb}$ is invariant under $w_\theta$.
\end{proof}%
\begin{proof}\emph{(Proof of Proposition~\ref{prop: rev_iss}.)}
For
	 $t \in (0, \norm{\targetdensity}_\infty)$ and $A,B \in \sigmaalgebra{\sphere{\dim -1}}$ Lemma~\ref{L: Liouville measure invariant under T} implies
	\begin{align*}
	& \area{\dim -2} \int_{B \cap \levelset{t}} H_t(x,A) \, \unisphere{\dim -1}(\d x)   \\
	& \quad 
	=   \int_{B \cap  \levelset{t}} \int _{\subsphere{\dim -2}{x}} \frac{1}{{\leb}(\glevelset{x}{v}{t})}\int_{\glevelset{x}{v}{t}} \mathbbm{1}_A\big(\geodesic{(x,v)}(\theta)\big) \, \d \theta\, \unitangsphere{x}(\d v)  \, \unisphere{\dim -1}(\d x) \\
	& \quad = \int_{[0, 2\uppi)} \int_{\sphere{\dim -1}} \int _{\subsphere{\dim -2}{x}}  \frac{\mathbbm{1}_{\glevelset{T_\theta(x}{v)}{t}}(\theta) \mathbbm{1}_{B \cap \levelset{t}}(\geodesic{(x,v)}(\theta)) \mathbbm{1}_A \big(\geodesic{T_\theta(x,v)}(\theta)\big)}{{\leb}(\glevelset{T_\theta(x}{v)}{t})}
	\, \unitangsphere{x}(\d v)  \unisphere{\dim -1}(\d x) \, \d \theta.
	\end{align*}
	Then \eqref{Eq: Level set identity}, Lemma \ref{L: T indentities geodesic} and Lemma \ref{L: T indentities - tau of level set under T} yield
	\begin{align*}
	& \area{\dim -2} \int_{B \cap \levelset{t}} H_t(x,A) \, \unisphere{\dim -1}(\d x)  \\
	& \quad = \int_{[0, 2\uppi)} \int_{\sphere{\dim -1}} \int _{\subsphere{\dim -2}{x}}  \frac{1}{{\leb}(\glevelset{x}{v}{t})}\mathbbm{1}_{\levelset{t}}(x) \mathbbm{1}_{B}(\geodesic{(x,v)}(\theta)) \mathbbm{1}_{\glevelset{x}{v}{t}}(\theta) \mathbbm{1}_A(x) 
	%		Daniel: Slight overfull box
	%		\\
	%		& \quad \hspace{9cm} \times\, 
	\unitangsphere{x}(\d v)  \unisphere{\dim -1}(\d x) \, \d \theta  \\
	& \quad= \area{\dim -2} \int_{A \cap \levelset{t}} H_t(x,B) \, \unisphere{\dim -1}(\d x).
	\end{align*}
	Hence, $H_t$ is reversible w.r.t. 
	$\mathcal{U}_{L(t)}=\frac{1}{\unisphere{\dim -1}(\levelset{t})}\,\unisphere{\dim -1}\vert_{\levelset{t}}$. 
	Lemma \ref{L: What to show for reversibility} then implies that $H$ is reversible with respect to $\target$.
\end{proof}

\section{Proof of Lemma~{\ref{L: Essential step for smallness of S^d}}}\label{Sec: Integral estimate for smallness}
In this section we prove the integral estimate of Lemma~\ref{L: Essential step for smallness of S^d}. 
A major part of deriving it consists of handling the measure 
resulting from exploring the sphere along the great circle passing through a fixed $x \in \sphere{\dim -1}$.
At first sight, one could think that this measure is $\unisphere{\dim -1}$.
Keeping in mind though that, interpreting $x$ as the ``north pole'', the great circles lie much ``denser'' at the poles than at the equator,
it becomes rather clear that $\unisphere{\dim -1}$ can only be a lower estimate for the measure obtained by exploring $\sphere{\dim -1}$ via the great circles through $x$.

To prove the desired
 statement, we make use of an expression for $\unisphere{\dim -1}$ in terms of polar coordinates \citep[see][Corollary 16.19]{Schilling}.
Define the polar coordinate transformations
\begin{align*}
  f_{\dim -1}: (0, \uppi)^{\dim-2} \times (-\uppi, \uppi)
\to& \mathbb{R}^{\dim} \setminus \{ (x_1,\dots,x_d)^T\in\mathbb{R}^d \mid x_{\dim} = 0, x_{\dim -1} \leq 0 \},\\
(\theta_1, \ldots, \theta_{\dim-1}) & \mapsto
\begin{pmatrix}
\cos(\theta_1)                                                               \\ \sin(\theta_1)\cos(\theta_2)\\  
\prod_{i=1}^{2}\sin(\theta_i)\cos(\theta_3) \\ \vdots \\
\prod_{i=1}^{\dim-2}\sin(\theta_{i})
\cos(\theta_{\dim -1}) \\
\prod_{i=1}^{\dim-1}\sin(\theta_i)
\end{pmatrix},
\end{align*}
and the absolute value of the determinant of their Jacobians
\[
J_{\dim -1}( \theta_1,\ldots,\theta_{\dim -1}) =  \sin^{\dim -2}(\theta_1)\sin^{\dim -3}(\theta_2)\cdot \ldots \cdot \sin(\theta_{\dim -2}).
\]
Then for all $A \in \sigmaalgebra{\sphere{\dim -1}}$ we have 
\begin{equation}\label{Eq: Expression for volume measure}
	\begin{split}
		\unisphere{\dim -1}(A)
	 = \int_0^\uppi \ldots \int_0^\uppi \int_{-\uppi}^\uppi	 \mathbbm{1}_{A}\big( f_{\dim -1}(\theta_1, \ldots, \theta_{\dim -1}) \big)
		J_{\dim -1}(\theta_1,\ldots,\theta_{\dim -1})  \, \d \theta_{\dim -1} \ldots \d \theta_1.
	\end{split}
\end{equation}

\begin{proof}\emph{(Proof of Lemma~\ref{L: Essential step for smallness of S^d}.)}
	Recall that $\unitangsphere{x} = \varphi_x(\unisphere{\dim -2})$ where $\varphi_x$ is defined in \eqref{Eq: Definition of alpha}. 
	This allows us to shift from $\subsphere{\dim -2}{x}$ to $\sphere{\dim -2}$, 
	where we may use the previous explicit expression in polar coordinates.
	That is,
	applying \eqref{Eq: Expression for volume measure} to $\unisphere{\dim -2}$ we get
	\begin{align*}
		& \int_0^\uppi \int_{\subsphere{\dim -2}{x}} \mathbbm{1}_{A}\big(\geodesic{(x,v)}(\theta)\big) \, \unitangsphere{x}(\d v) \, \d \theta
		= \int_0^\uppi \int_{\sphere{\dim -2}} \mathbbm{1}_{A}\big(\geodesic{(x,\varphi_x(y))}(\theta)\big) \, \unisphere{\dim -2}(\d y) \, \d \theta \\
		& \qquad = \int_0^\uppi \int_0^\uppi \ldots \int_0^\uppi \int_{-\uppi}^\uppi
		\mathbbm{1}_{A}\left(\geodesic{\left(x,\varphi_x\left( f_{\dim -2}(\theta_1, \ldots, \theta_{\dim -2}) \right)\right)}(\theta)\right)
		J_{\dim-2}(\theta_1, \ldots, \theta_{\dim-2})                                                                                               
		 \d \theta_{\dim-2} \ldots \d \theta_1 \,\d \theta.
	\end{align*}
	Now, we can use the outer integral (corresponding to travelling along the great circles) to add a dimension in the explicit polar coordinate representation of the volume measure on the sphere at the cost of introducing a correction term. To this end,
	extend the map $\varphi_x$ to
	\begin{align*}
		\widetilde{\varphi}_x: \sphere{\dim-1} & \to \sphere{\dim-1},\qquad
		(y_1, \ldots, y_\dim) \mapsto y_1 x + \sum_{i = 1}^{\dim-1} y_{i+1} e_i^{(x)}.
	\end{align*}
	Observe that this reparametrization of $\sphere{\dim -1}$ is compatible with the geodesic structure of the sphere,
	because it respects the basis of $\subsphere{\dim -2}{x}$ chosen by the map $\varphi_x$.
	More precisely,
	\begin{align*}
		\geodesic{\left(x,\varphi_x\left( f_{\dim -2}(\theta_1, \ldots, \theta_{\dim -2}) \right)\right)}(\theta)
		= \widetilde{\varphi}_x\big( f_{\dim -1}(\theta, \theta_1, \ldots, \theta_{\dim-2})\big).
	\end{align*}
	Furthermore, defining 
	\begin{align*}
		g: \mathbb{R}^{\dim } \setminus \big\{ (y_1,\dots,y_d)\in\mathbb{R}^d \mid |y_1| > 1\big \} & \to \mathbb{R}_+, \qquad
		(y_1, \ldots, y_{\dim }) \mapsto \frac{1}{\sin^{\dim -2}(\arccos(y_1))},
	\end{align*}
	we have
	\begin{align*}
		J_{\dim-2}(\theta_1, \ldots, \theta_{\dim-2})
		= g\big(f_{\dim -1}(\theta, \theta_1, \ldots, \theta_{\dim-2})\big) J_{\dim-1}( \theta,\theta_1, \ldots, \theta_{\dim-2}).
	\end{align*}
	Therefore
	\begin{align*}
		& \int_0^\uppi \int_{\subsphere{\dim -2}{x}} \mathbbm{1}_{A}\big(\geodesic{(x,v)}(\theta)\big) \, \unitangsphere{x}(\d v) \, \d \theta \\
		&\qquad = \int_0^\uppi \int_0^\uppi \ldots \int_0^\uppi \int_{-\uppi}^\uppi
		\mathbbm{1}_{A}\left(\widetilde{\varphi}_x\big( f_{\dim -1}(\theta, \theta_1, \ldots, \theta_{\dim-2})\big)\right)
		g\big(f_{\dim -1}(\theta, \theta_1, \ldots, \theta_{\dim-2})\big)                                                                     \\
		&\qquad \hspace{6cm}	\cdot J_{\dim-1}( \theta,\theta_1, \ldots, \theta_{\dim-2})
		\, \d \theta_{\dim-2} \ldots \d \theta_1 \,\d \theta.
	\end{align*}
	Applying \eqref{Eq: Expression for volume measure} for $\unisphere{\dim -1}$ we obtain
	\begin{align*}
		& \int_0^\uppi \int_{\subsphere{\dim -2}{x}} \mathbbm{1}_{A}\big(\geodesic{(x,v)}(\theta)\big) \, \unitangsphere{x}(\d v) \, \d \theta
		= \int_{\sphere{\dim -1}}\mathbbm{1}_{A}\big(\widetilde{\varphi}_x(y)\big) g(y)\, \unisphere{\dim -1}(\d y).
	\end{align*}
	Observe that $0 \leq \sin^{\dim-2} \circ \arccos \leq 1$, such that $g \geq 1$. Moreover, $\unisphere{\dim -1}$ is invariant under the orthogonal map $\widetilde{\varphi}_x$. Hence,
	\begin{align*}
		& \int_0^\uppi \int_{\subsphere{\dim -2}{x}} \mathbbm{1}_{A}\big(\geodesic{(x,v)}(\theta)\big) \, \unitangsphere{x}(\d v) \, \d \theta
		\geq \int_{\sphere{\dim-1}} \mathbbm{1}_{A}\big(\widetilde{\varphi}_x(y)\big) \, \unisphere{\dim -1}(\d y)
		=  \unisphere{\dim -1}(A).
	\end{align*}
	Finally, take into account that $\mathcal{U}_{\sphere{\dim-1}} = \frac{1}{\area{\dim -1}}\, \unisphere{\dim -1}$, $\mathcal{U}_{\subsphere{\dim -2}{x}}=\frac{1}{\area{\dim -2}}\unitangsphere{x}$ and $\frac{\area{\dim -1}}{\area{\dim -2}}\geq \frac{\sqrt{2\uppi}}{\sqrt{d-1}}$, cf. \citep[Lemma 6]{MatheNovak} as well as the formulas for $\area{\dim-1}, \area{\dim-2}$ in terms of the Gamma function.
\end{proof}
\section{Proof of Proposition~\ref{L: Reversibility SSS}}\label{Sec: Reversibility of SSS}
\citet{ReversibilityEllipticalSliceSampler} provide a formal description, with a reversibility result, of the shrinkage scheme that we use in the proof of Proposition~\ref{L: Reversibility SSS}. For convenience of the reader we restate 
it
 and state the corresponding reversibility result. 

To this end we call a set $S \in \sigmaalgebra{[0, 2\uppi)}$ \emph{open on the circle} if for all $\theta \in S$ there exists $\varepsilon > 0$ such that $\{a \mod 2\uppi \mid |a-\theta| < \varepsilon, a \in \mathbb{R} \} \subseteq S$.
Moreover, for $a, b \in [0, 2\uppi)$ define generalized intervals
\[
I(a,b) := \begin{dcases} [0, b) \cup [a, 2\uppi), & a > b \\
[a, b), & a < b\\
[0, 2\uppi), & a = b,
\end{dcases} \qquad 
\bar{I}(a,b) := \begin{dcases} [0, b) \cup [a, 2\uppi), & a > b \\
[a, b), & a < b\\
\emptyset, & a = b,
\end{dcases}
\]
that appear in Algorithm~\ref{A: SP}, which is called as $\overline{\text{shrink}}(\theta, S)$ for $S \in \sigmaalgebra{[0, 2\uppi)}$ being open on the circle and $\theta \in S$.
\begin{algorithm}
	\caption{Algorithm 2.2 from \citet{ReversibilityEllipticalSliceSampler} with input $S \in \sigmaalgebra{[0, 2\uppi)}$ and $\theta\in S$, called by $\overline{\text{shrink}}(\theta, S)$.}
	\algorithmicrequire \ current state $\theta \in S$ \label{A: SP}\\
	\algorithmicensure \ step-size $\alpha$
	\begin{algorithmic}[1]
		\STATE Set $i := 1$ and draw $\Lambda_i \sim \uniform{0, 2\uppi}$, call the result $a_i$.
		\STATE Set $a^{\min}_i := a_i$ and $a_i^{\max} := a_i$.
		\WHILE{$a_i \notin S$}
		\IF{$ a_i \in \bar{I}(a^{\min}_i , \theta)$}
		\STATE Set $a^{\min}_{i+1} : = a_i$ and $a^{\max}_{i+1} : = a^{\max}_{i}$.
		\ELSE
		\STATE Set $a^{\min}_{i+1} = a^{\min}_{i}$ and $a^{\max}_{i+1} = a_i$.
		\ENDIF
		\STATE Draw $\Lambda_{i+1} \sim \mathcal{U}_{I(a^{\min}_{i+1}, a^{\max}_{i+1})}$, call the result $a_{i+1}$.
		\STATE Set $i:= i+1$.
		\ENDWHILE
		\STATE Return $\alpha:=a_i$.
	\end{algorithmic}
\end{algorithm}

For a given set $S \in \sigmaalgebra{[0, 2\uppi)}$ that is open on the circle the transition kernel defined by Algorithm~\ref{A: SP} is denoted by
\begin{equation} \label{Eq: Definition shrinkage kernel}
	\bar{Q}_S : S \times \sigmaalgebra{S} \to [0,1],
\end{equation}
that is,
\[
\bar{Q}_S(\theta, A) = \mathbb{P}\big(\overline{\text{shrink}}(\theta, S) \in A \big), \qquad \forall \theta \in [0, 2\uppi), A \in \sigmaalgebra{[0, 2\uppi)}.
\]
Similarly as in \citet{ReversibilityEllipticalSliceSampler} observe that for $S = \glevelset{x}{v}{t}$ with $x \in \sphere{\dim -1}$, $v \in \subsphere{\dim -2}{x}$ and $t \in (0,p(x))$, holds
\begin{equation}\label{Eq: Shrinkage distribution in terms of shrinkage kernel}
	\begin{split}
		\shrinkagedist{x}{v}{t}(\cdot) &=\mathbb{P}\big(\shrink{x}{v}{t} \in \cdot \mod 2 \uppi \big)
		=  \mathbb{P}\big(\overline{\text{shrink}}(0, \glevelset{x}{v}{t}) \in \cdot \big)\\
		 &= \bar{Q}_{\glevelset{x}{v}{t}}(0, \cdot),
	\end{split}
\end{equation}
where the distribution $\shrinkagedist{x}{v}{t}$ is provided in Definition~\ref{D: Definition of shrinkage distribution}. We state two useful properties of the kernel of the shrinkage procedure $\bar{Q}_S$ that are proven by {\citet{ReversibilityEllipticalSliceSampler}}.

\begin{lemma}[{\citealp[Theorem 2.10]{ReversibilityEllipticalSliceSampler}}]\label{L: Reversibility of shrinkage procedure}
	Let $S \in \sigmaalgebra{[0, 2\uppi)}$ be non-empty and open on the circle. Then, $\bar{Q}_S$ defined in \eqref{Eq: Definition shrinkage kernel} is reversible w.r.t. $\mathcal{U}_S$.
\end{lemma}

\begin{lemma}[{\citealp[Lemma 2.12]{ReversibilityEllipticalSliceSampler}}]\label{L: shrinkage procedure under pushforward}
	Let $S \in \sigmaalgebra{[0, 2\uppi)}$ be non-empty and open on the circle, $\theta \in [0, 2\uppi)$, and define
	$
		g_\theta	: [0, 2\uppi) \to [0, 2\uppi)$  
		as
		$a \mapsto (\theta - a) \ \text{mod} \ 2 \uppi.
	$
	Then
	\[
		\bar{Q}_{g_\theta^{-1}(S)}\left({g_\theta^{-1}(\alpha)}, g_\theta^{-1}(A) \right)	= \bar{Q}_S({\alpha}, A), \qquad \forall {\alpha \in S}, A \in \sigmaalgebra{S}.
	\]
\end{lemma}
Intuitively, the map $g_\theta$ from the previous lemma corresponds to following a line in reverse direction with an offset of $\theta$.
The lemma tells us if we apply this motion to all inputs of the shrinkage procedure simultaneously, its effects cancel.
We formulate the consequences for our setting.

By Lemma~\ref{L: T indentities geodesic} and the $2\uppi$-periodicity of sine and cosine, we have
\[
\geodesic{(x,y)}\left( g_\theta(a) \right) = \geodesic{T_\theta(x,v)}(a), \qquad \forall \ x \in \sphere{\dim -1}, v \in \subsphere{\dim -2}{x}, \theta,a \in [0, 2\uppi).
\]
This implies for all $x \in \sphere{\dim -1}, v \in \subsphere{\dim -2}{x}, \theta \in [0, 2\uppi), t\in(0,p(x))$ and  $B \in \sigmaalgebra{\sphere{\dim -1}}$ that
\[
g_\theta^{-1}\big(\glevelset{x}{v}{t} \big) = \glevelset{T_\theta(x}{v)}{t}
\qquad\text{and}\qquad g_\theta^{-1}\left( \geodesic{(x,v)}^{-1}(B) \right) = \geodesic{T_\theta(x,v)}^{-1}(B).
\]
Hence, by applying Lemma \ref{L: shrinkage procedure under pushforward} for $A = \geodesic{(x,v)}^{-1}(B)$, $\alpha = \theta$ and $S = \glevelset{x}{v}{t}$, we obtain
\begin{equation}\label{Eq: shrinkage procedure with level set under T_theta}
	\begin{split}
		\bar{Q}_{\glevelset{T_\theta (x}{v)}{t}}\left(0, \geodesic{T_\theta(x,v)}^{-1}(B)\right)
		= \bar{Q}_{g_\theta^{-1}\big(\glevelset{x}{v}{t} \big)}\left(0,g_\theta^{-1}\left( \geodesic{(x,v)}^{-1}(B) \right) \right)
		= \bar{Q}_{\glevelset{x}{v}{t}}(\theta,\geodesic{(x,v)}^{-1}(B)).
	\end{split}
\end{equation}
To prove Proposition~\ref{L: Reversibility SSS} we show reversibility of $\widetilde{H}_t$ w.r.t. 
	$\mathcal{U}_{L(t)}$
	for all $t \in (0, \norm{p})_{\infty}$ and then conclude the assertion by Lemma \ref{L: What to show for reversibility}.
	To achieve this, we use the invariance of the Liouville measure under $T_\theta$ (which corresponds to ``a forward move with a U-turn'').
	This boils down to starting the shrinkage procedure at a random point on the geodesic level set.

\vspace{0.5ex}
\begin{proof}\emph{(Proof of Proposition~\ref{L: Reversibility SSS}.)}
	Let $t \in (0,  \norm{\targetdensity}_\infty)$ and $A,B \in \sigmaalgebra{\sphere{\dim -1}}$. Due to the lower semicontinuity of $\targetdensity$, the level set $\levelset{t}$ is open. Thus, $\glevelset{x}{v}{t}$ is non-empty and open on the circle for all $x \in \levelset{t}$ with $v \in \subsphere{\dim -2}{x}$. 
Moreover, Lemma~\ref{lem: low_semi_implies_geo_lev_>0} implies that $\leb( \glevelset{x}{v}{t} ) > 0$ for all $ x \in \levelset{t}$ and $v \in \subsphere{\dim -2}{x}$.
	%%%%%%%%%%%%%%%%%%%%%%%
	% Replace the previous argument using lower semicontinuity by the following argument to obtain a proof that does not use lower semicontinuity.
	%%%%%%%%%%%%%%%%%%%%%%%%
	%	Observe that for all $x \in L(t) \setminus N(t)$ holds
	%	\[
	%		\unitangsphere{x}\left( V(x,t)^C \right) = 0	
	%	\]
	%	by definition of $N(t)$, see Lemma \ref{L: Auxilary statement for well definedness}.
	%	Moreover, by definition of $V(x,t)$, see Lemma \ref{L: A.s. statements about tau(L(x,v,t)) > 0}, we have
	%	\[
	%		 \{(x,v) \in \tangentbundle \mid x \in \levelset{t} \text{ and } {\color{blue}\leb} (\glevelset{x}{v}{t}) = 0 \} =
	%		 \bigcup_{x \in \levelset{t}} \{x\} \times V(x,t)^C.
	%	\]
	%	Therefore
	%	\begin{equation*}
		%		\begin{split}
			%			&\liouville\left( \{(x,v) \in \tangentbundle \mid x \in \levelset{t} \text{ and } {\color{blue}\leb} (\glevelset{x}{v}{t}) = 0 \} \right)\\
			%			&\qquad= \liouville\left( \bigcup_{x \in \levelset{t}} \{x\} \times V(x,t)^C \right)
			%			= \int_{\levelset{t}} \unitangsphere{x}\left( V(x,t)^C \right) \, \unisphere{\dim -1}(\d x)\\
			%			& \qquad= \int_{L(t) \setminus N(t)} \unitangsphere{x}\left( V(x,t)^C \right) \, \unisphere{\dim -1}(\d x)
			%			= 0
			%		\end{split}
		%	\end{equation*}
	%	where we used that $\unisphere{\dim -1} (N(t)) = 0$ by Lemma \ref{L: Auxilary statement for well definedness}.
	%
	%%%%%%%%%%%%%%%%%%%%%%%%%%%
	%%%%%%%%%%%%%%%%%%%%%%%%%%	
	Hence, we have, exploiting \eqref{Eq: Shrinkage distribution in terms of shrinkage kernel}, that
	\begin{align*}
		 &\area{\dim -2} \int_{B \cap \levelset{t}} \widetilde{H}_t(x,A) \, \unisphere{\dim -1}(\d x)                                                                           
		  = \int_{B \cap L(t)} \int_{\subsphere{\dim -2}{x}} \bar{Q}_{\glevelset{x}{v}{t}}(0, \geodesic{(x,v)}^{-1}(A))\, \unitangsphere{x}(\d v) \, \unisphere{\dim -1}(\d x) \\
		& \qquad  = \int_{[0,2\uppi)} \int_{B \cap \levelset{t}} \int_{\subsphere{\dim -2}{x}} \frac{\mathbbm{1}_{\glevelset{x}{v}{t}}(\theta)}{{\leb}(\glevelset{x}{v}{t})}
		\bar{Q}_{\glevelset{x}{v}{t}}(0, \geodesic{(x,v)}^{-1}(A))                                                                                                                     
		\unitangsphere{x}(\d v) \, \unisphere{\dim -1}(\d x)\, \d \theta.
	\end{align*}
	Then, Lemma~\ref{L: Liouville measure invariant under T} yields
	\begin{align*}
		&\area{\dim -2} \int_{B \cap \levelset{t}} \widetilde{H}_t(x,A) \, \unisphere{\dim -1}(\d x)                                                            
		 = \int_{[0,2\uppi)} \int_{\sphere{\dim -1}} \int_{\subsphere{\dim -2}{x}}\mathbbm{1}_{B \cap \levelset{t}}\left( \geodesic{(x,v)}(\theta) \right)
		\frac{\mathbbm{1}_{\glevelset{T_\theta(x}{v)}{t}}(\theta)}{{\leb}(\glevelset{T_\theta (x}{v)}{t})}                                                       \\
		& \quad \hspace{6cm} \times	\bar{Q}_{\glevelset{T_\theta(x}{v)}{t}}(0, \geodesic{T_\theta(x,v)}^{-1}(A)) \
		\unitangsphere{x}(\d v) \, \unisphere{\dim -1}(\d x)\, \d \theta.
	\end{align*}
	Using \eqref{Eq: shrinkage procedure with level set under T_theta}, \eqref{Eq: Level set identity}, Lemma~\ref{L: T indentities geodesic} and Lemma \ref{L: T indentities - tau of level set under T}, we obtain
	\begin{align*}
		& \area{\dim -2} \int_{B \cap \levelset{t}} \widetilde{H}_t(x,A) \, \unisphere{\dim -1}(\d x)                                                                                                                                     \\
		&= \int_{[0,2\uppi)} \int_{\sphere{\dim -1}} \int_{\subsphere{\dim -2}{x}} \mathbbm{1}_{B}(\geodesic{(x,v)}(\theta)) \frac{ \mathbbm{1}_{\glevelset{x}{v}{t}}(\theta) \mathbbm{1}_{\levelset{t}}(x) }{{\leb}(\glevelset{x}{v}{t})}
		\bar{Q}_{\glevelset{x}{v}{t}}(\theta, \geodesic{(x,v)}^{-1}(A))  
%		Daniel: Overfull box  
%		 \\
%		& \hspace{9,5cm}\times
		\unitangsphere{x}(\d v)  \unisphere{\dim -1}(\d x) \d \theta                                                                                                                                                  \\
		&= \int_{\levelset{t}} \int_{\subsphere{\dim -2}{x}} \frac{1}{{\leb}(\glevelset{x}{v}{t})} \int_{\geodesic{(x,v)}^{-1}(B)\cap \glevelset{x}{v}{t}}
		\bar{Q}_{\glevelset{x}{v}{t}}(\theta, \geodesic{(x,v)}^{-1}(A))                                                                                                                                                                           \d \theta \, \unitangsphere{x}(\d v)\, \unisphere{\dim -1}(\d x).
	\end{align*}
	As  $\glevelset{x}{v}{t}$ is open on the circle and non-empty for all $x \in \levelset{t}$ and $v \in \subsphere{\dim -2}{x}$, we may apply Lemma~\ref{L: Reversibility of shrinkage procedure} and get
	\begin{align*}
		& \area{\dim -2} \int_{B \cap \levelset{t}} \widetilde{H}_t(x,A) \, \unisphere{\dim -1}(\d x)                                                              \\
		& \qquad = \int_{\levelset{t}} \int_{\subsphere{\dim -2}{x}} \frac{1}{{\leb}(\glevelset{x}{v}{t})} \int_{\geodesic{(x,v)}^{-1}(A)\cap \glevelset{x}{v}{t}}
		\bar{Q}_{\glevelset{x}{v}{t}}(\theta, \geodesic{(x,v)}^{-1}(B))                                                                                                   
	 \d \theta \, \unitangsphere{x}(\d v)\, \unisphere{\dim -1}(\d x).
	\end{align*}
	Performing the same arguments in reversed order, we get
	\[
	\area{\dim -2} \int_{B \cap \levelset{t}} \widetilde{H}_t(x,A) \, \unisphere{\dim -1}(\d x)
	=  \area{\dim -2} \int_{A \cap \levelset{t}} \widetilde{H}_t(x,B) \, \unisphere{\dim -1}(\d x).
	\]
	Thus, by Lemma \ref{L: What to show for reversibility}, we obtain reversibility of $\widetilde{H}$ with respect to $\target$.
\end{proof}
\vspace*{-5ex}
\section{Random-walk Metropolis-Hastings on the Sphere}\label{app:rwmh}
The random-walk Metropolis-Hastings (RWMH) algorithm uses an isotropic Gaussian proposal kernel. As suggested by \citet{DimIndependentMCMCOnSpheres}, for current state $x\in\sphere{\dim-1}$ we choose an auxiliary point $\sqrt{r} x $ in the ambient space $\reals{\dim}$ by generating a radius $\sqrt{r}$, with $r$ being a realization of $R \sim {\rm Gamma}(d/2, 1/2)\, .$
%
%\[
%R \sim {\rm Gamma}(d/2, 1/2)\, .
%\]
%
Then, given $\sqrt{r} x$ we sample a realization $y$ from $Y \sim {\rm Normal}\left(\sqrt{r}x, \varepsilon^2 \mathrm{Id}_\dim\right),$
%\[
%Y \sim {\rm Normal}\left(rx, \varepsilon^2 \mathrm{Id}_\dim\right),
%\]
where $\varepsilon$ is the step-size of the random walk. Since this $y$ does not yet lie on the sphere, we radially project and propose $y/\Vert y \Vert$, which finally is accepted or rejected using the usual acceptance ratio. In Algorithm~\ref{alg:rwmh} we provide the corresponding pseudocode. Note that we also introduce mixture-MH, which can be interpreted as a variant that combines RWMH with  independent Metropolis. For details we refer to Algorithm~\ref{alg:kernel-mixture-sampler}, which is specifically employed for testing the mixture of vMF distributions.
%\vspace*{-4ex}
%
\begin{algorithm}[htb]
	\caption{Reprojected RWMH on $\sphere{\dim-1}$ for step-size $\varepsilon > 0$.}
	\label{alg:rwmh}
	\algorithmicrequire current state $x \in \sphere{\dim-1}$\\
	\algorithmicensure next state $x'$
	\begin{algorithmic}[1]
		%     \INPUT current state $x \in \sphere{\dim-1}$, step-size $\epsilon > 0$
		%    \OUTPUT next state $x'$
		%
		\STATE Draw $R \sim {\rm Gamma}\left(\dim/2, 1/2\right)$, call the result $r$.
		\STATE Draw $Y \sim {\rm Normal}\left(\sqrt{r}\,x, \varepsilon^2 \mathrm{Id}_\dim \right)$, call the result $y$.
		\STATE Set $z = y/\|y\|$.
		\STATE Draw $U \sim \mathcal{U}_{(0, 1)}$, call the result $u$.
		\IF{$u \le \min\left\{1, p(z) / p(x) \right\}$}
		\STATE $x' = z$
		\ELSE
		\STATE $x' = x$
		\ENDIF
	\end{algorithmic}
\end{algorithm}
\begin{algorithm}[htb]
	\caption{Mixture of RWMH and Independence Sampler on $\sphere{\dim-1}$ with mixture hyperparameter $\alpha \in [0,1]$ and step-size $\varepsilon > 0$.}
	\label{alg:kernel-mixture-sampler}
	\algorithmicrequire current state $x \in \sphere{\dim-1}$\\
	\algorithmicensure next state $x'$
	\begin{algorithmic}[1]
		\STATE Draw $U_1 \sim \mathcal{U}_{(0, 1)}$, call the result $u_1$.
		\IF{$u_1 < \alpha$}
		\STATE // \texttt{Use random walk proposal}
		\STATE Draw $R \sim {\rm Gamma}\left(\dim/2, 1/2\right)$, call the result $r$.
		\STATE Draw $Y \sim {\rm Normal}\left(\sqrt{r}\,x, \varepsilon^2 \mathrm{Id}_\dim \right)$, call the result $y$.
		\STATE Set $z = y/\|y\|$.
		\ELSE
		\STATE // \texttt{Use independent proposal}
		\STATE Draw $Y \sim {\rm Normal}\left(0, \mathrm{Id}_\dim \right)$, call the result $y$.
		\STATE Set $z = y/\|y\|$.
		\ENDIF
		\STATE Draw $U_2 \sim \mathcal{U}_{(0, 1)}$, call the result $u_2$.
		\IF{$u_2 \le \min\left\{1, p(z) / p(x) \right\}$}
		\STATE $x' = z$
		\ELSE
		\STATE $x' = x$
		\ENDIF
	\end{algorithmic}
\end{algorithm}
\section{Hamiltonian Monte Carlo on the Sphere}\label{app:hmc}
Since RWMH suffers from diffusive behavior, we also test spherical Hamiltonian Monte Carlo (HMC) suggested by \citet{Lan14} as an alternative MCMC approach. In spherical HMC, the sample space is first augmented by {\em momenta} or {\em velocities} $v \in \reals{\dim}$ that are in the the tangent space $E_x$ to $\sphere{\dim-1}$ at the current sample $x$. The momenta follow a standard Normal distribution, and the Markov chain is generated in the space $\{(x,v) \mid x \in \sphere{\dim -1}, v \in E_x\}$. Diffusive behavior is suppressed by the special type of proposal step that solves Hamilton's equations of motion by using a leapfrog integrator \citep{Neal11}. During leapfrog integration, the gradient of the log probability $\nabla \log p(x)$ guides the Markov chain, thereby reaching nearby modes in much shorter time than RWMH. The spherical HMC algorithm is detailed in Algorithm \ref{alg:hmc}. In contrast to standard HMC, spherical HMC involves a rotation of the positions and momenta during leapfrog integration where the rotation matrix is a Givens rotation $G(v/\|v\|, x, \theta)$ in the plane spanned by the momentum $v$ and the position $x$ (see lines 8--10 in Algorithm \ref{alg:hmc} and Equation \ref{Eq: Givens rotation}). In addition to the step-size parameter $\varepsilon>0$, we also need to choose the number of leapfrog steps $T \in \mathbb N$. In our experiments, we always set $T=10$.\\[-6ex]

\section{Step-size Tuning for RWMH and HMC}\label{Sec: Step-size tuning}
Both RWMH and spherical HMC involve a step-size parameter $\varepsilon$. Because a good choice of $\varepsilon$ depends on the particular shape of our target distribution, we first find $\varepsilon$ automatically during a burn-in phase. During burn-in, we increase $\varepsilon$ by a factor 1.02, if the proposal (based on the current value of $\varepsilon$) is accepted. We decrease $\varepsilon$ by a factor 0.98 if the proposed state is rejected. After a burn-in phase, the value of $\varepsilon$ is kept fixed.

\begin{algorithm}[htb]
	\caption{Hamiltonian Monte Carlo on $\sphere{\dim-1}$ for step-size $\varepsilon > 0$ and number of integration steps $T$. }
	\label{alg:hmc}
	\algorithmicrequire current state $x \in \sphere{\dim-1}$ \\
	\algorithmicensure  next state $x'$
	\begin{algorithmic}[1]
		\STATE Draw $V \sim {\rm Normal}(0, \mathrm{Id}_\dim)$, call the result $v$.
		\STATE Set $v_1 = (\mathrm{Id}_\dim - xx^T)\, v$, $x_1=x$ and $t=1$.
%		.
%		\STATE Set $x_1 = x$.
%		\STATE Set $t = 1$.
		\WHILE{$t \le T$}
		\STATE Set $v_{t + 1/2} = v_t + \frac{\varepsilon}{2}\, \left(\mathrm{Id}_\dim - x_t x_t^T\right)\, \nabla \log p(x_t)$.
		\STATE Set $R = \givens{v_{t + 1/2}/\|v_{t + 1/2}\|}{x_t}{\varepsilon\,\|v_{t + 1/2}\|}$.
		\STATE Set $x_{t + 1} = R\, x_t$.
		\STATE Set $v_{t + 1} = R\, v_{t + 1/2} + \frac{\varepsilon}{2}\, \left(\mathrm{Id}_\dim - x_{t+1}x_{t+1}^T\right)\, \nabla \log p(x_{t+1})$.
		\STATE Set $t:=t+1$.
		%    Increment $t$ by 1.
		\ENDWHILE
		\STATE Draw $U \sim \mathcal{U}_{(0, 1)}$, call the result $u$.
		\IF{$u < \min\left\{1, \exp\left(\|v_1\|^2/2 - \|v_{T+1}\|^2/2 \right)\, {p(x_{T+1})}/{p(x)} \right\}$}
		\STATE Set $x' = x_{T+1}$.
		\ELSE
		\STATE Set $x' = x$.
		\ENDIF
	\end{algorithmic}
%\vspace*{-0.5ex}
\end{algorithm}
%\clearpage
\newcommand{\hide}[1]{{}}
\hide{\section{Tests on 50-dimensional Bingham Distribution.}
Tests for the 50-dimensional Bingham distribution with a concentration parameter $\kappa_{50} = 300$.
\begin{figure}[H]
	\centering{
		\includegraphics[width=0.9\textwidth]{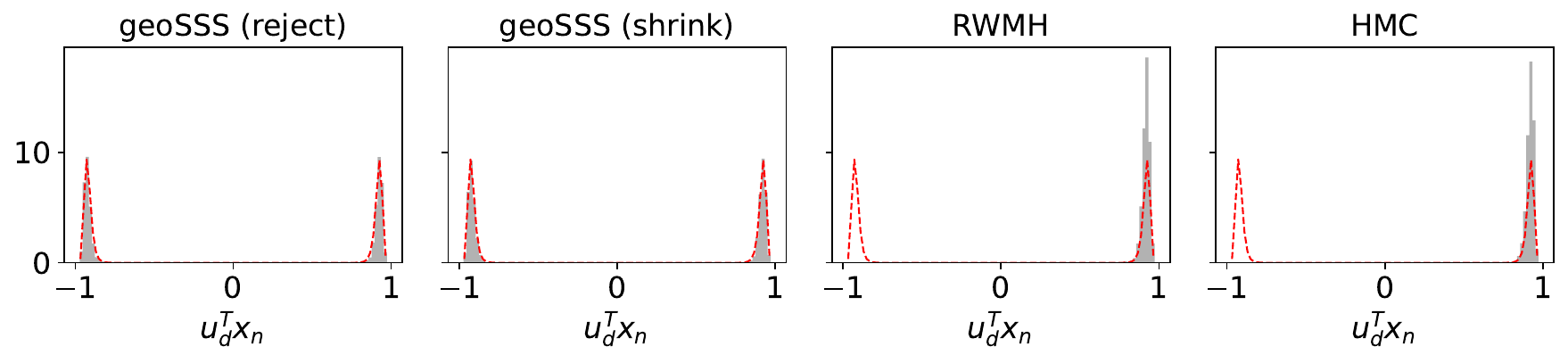}
	}
	\caption{Histograms of approximate Bingham samples ($\dim=50, \kappa_d=300$) projected on the first mode obtained with each MCMC method are shown in gray. The red dashed line indicates the baseline obtained with the acceptance/rejection sampler of \citet{Kent18}. }\label{fig:bing2-hist}
\end{figure}
\begin{figure}[H]
	\centering{
		\includegraphics[width=0.9\textwidth]{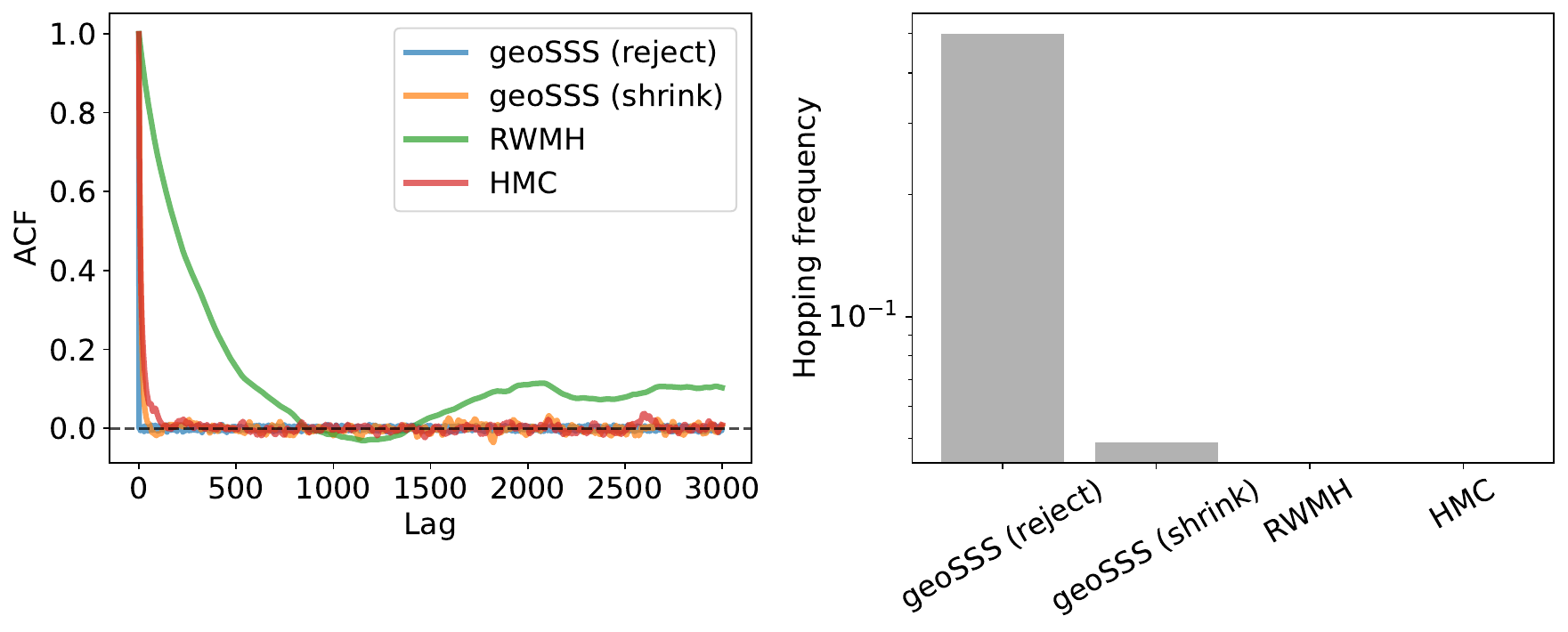}
	}
	\vspace*{-3ex}
	\caption{Left: Autocorrelation analysis of $x\mapsto u_{\dim}^Tx$ with respect to approximate samples from a Bingham distribution with $\dim=50$ and $\kappa_{\dim} = 300$. Right: Estimated hopping frequency between both modes of the Bingham distribution. Note that the hopping frequency is shown on a logarithmic axis. }\label{fig:bing2-acf}
	\vspace*{-5ex}
\end{figure}

\begin{figure}[H]
	\centering{
		\includegraphics[width=0.9\textwidth]{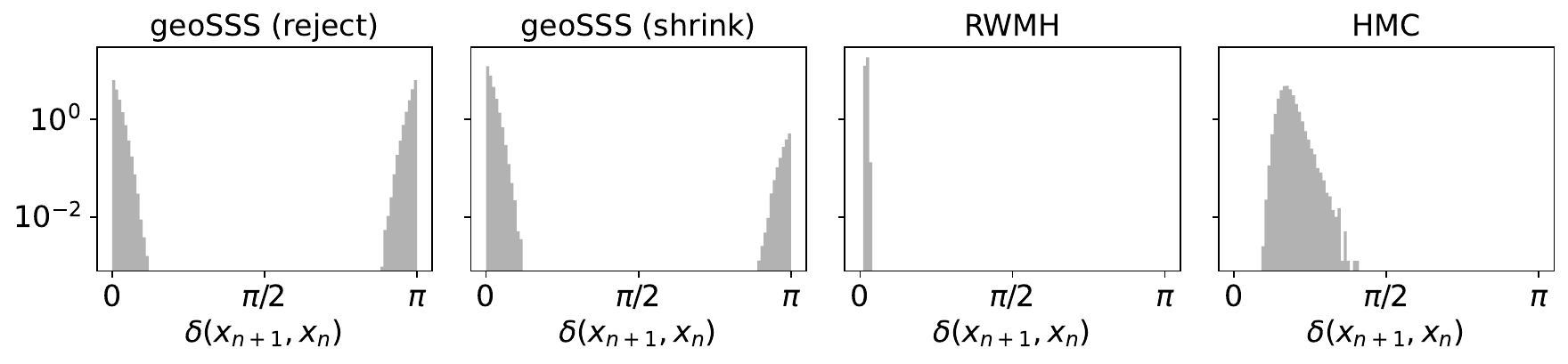}
	}
\vspace*{-3ex}
	\caption{Geodesic distance (log-scale) between successive approximate samples for a Bingham distribution with $\dim=50$ and $\kappa_d=300$. }\label{fig:bing2-dist}
\vspace*{-5ex}
\end{figure}
}

\section{Run Times of Numerical Illustrations} \label{supp-run-times}
We summarize the run times in seconds (s) for all the numerical experiments tested on Intel i7-10510U CPU in Table \ref{table:run-times}. In addition to the tests reported in Section \ref{Sec: Numerical experiments}, we provide a Jupyter notebook for testing the geodesic slice samplers on Bingham targets in various dimensions.\footnote{\href{https://github.com/microscopic-image-analysis/geosss/blob/main/scripts/Bingham.ipynb}{\ttfamily github.com/microscopic-image-analysis/geosss/scripts/Bingham.ipynb}} We also report run times for these targets. For brevity we name these experiments ``Bingham-d10" for the Bingham distribution with $d=10$, $\kappa_d = 30$ and $N = 10^5$; 
``Bingham-d50" for the Bingham distribution with $d=50$, $\kappa_d = 300$ and $N = 10^5$;
``vMF-d10" for the mixture of vMF distribution with $d=10$, $\kappa = 100$, $K=5$ and $N = 10^6$;
``curved-vMF-d3" for the curve vMF distribution with $d=3$, $\kappa = 500$ and $N = 10^6$;
``curved-vMF-d5" for curve vMF with $d=5$, $\kappa=800$ and $N=10^6$.

\begin{table}[tb]
	\centering
	\begin{threeparttable}
		\renewcommand{\arraystretch}{1.2}
		\begin{tabular}{lcccc}
			\toprule
			\rowcolor{orange!20}
			\textbf{Experiments} & \textbf{geoSSS (reject)} & \textbf{geoSSS (shrink)} & \textbf{RWMH} & \textbf{HMC} \\
			\midrule
			\rowcolor{orange!6} Bingham-d10      & 13.9 s & 8.5 s   & 3.5 s   & 28.4 s \\
			\rowcolor{orange!6} Bingham-d50      & 36.5 s & 13.1 s  & 3.5 s   & 32.0 s \\
			\rowcolor{orange!6} vMF-d10\tnote{*} & 2601.6 s & 812.9 s & 274.4 s & 1898.9 s \\
			\rowcolor{orange!6} curve-vMF-d3     & 17106.9 s & 5583.2 s & 1841.0 s & 12292.9 s \\
			\rowcolor{orange!6} curve-vMF-d5     & 42483.7 s & 5509.0 s & 1355.7 s & 8659.0 s \\
			\bottomrule
		\end{tabular}
		\begin{tablenotes}
			\small
			\item[*] The extensively tuned mixture-MH sampler employed only for this test took 577.0 s.
		\end{tablenotes}
		\caption{Comparison of run times for each MCMC method across different numerical experiments.}
		\label{table:run-times}
	\end{threeparttable}
\end{table}

\hide{\subsection{Bingham Distribution}
In our first test, we aim to sample from the Bingham distribution \citep{Mardia09} whose unnormalized density is defined as
\begin{equation}\label{eq:bingham}
	\bing{}(x) = \exp(x^T\!Ax)
\end{equation}
for $x\in\sphere{\dim-1}$. Without loss of generality, the matrix $A \in \reals{\dim\times\dim}$ is symmetric such that we can transform the variables into the eigenbasis of $A$. In the following, we assume that the variables underwent this transformation by letting $A = \text{diag}(\kappa_1, \ldots, \kappa_{\dim})$ where $\kappa_1 \le \kappa_2 \le \ldots \le \kappa_d$ are the eigenvalues of $A$. Because the Bingham distribution is invariant under shifts of the diagonal of $A$, i.e., invariant under $A\mapsto A + c\, \mathrm{Id}_d$, where $c\in\reals{}$ is some constant, we can let $\kappa_1 = 0$. The Bingham distribution is bimodal with symmetric modes at $\pm u_{\dim}$ where $u_{\dim} \in \sphere{\dim-1}$ is the eigenvector of $A$ with the largest eigenvalue $\kappa_{\dim}$. The maximal value of the logarithm of the unnormalized probability density is $\kappa_d$, i.e., $\max_{x\in\sphere{d-1}} \log \bing{}(x) = \kappa_{\dim}$.
Various algorithms for simulating the Bingham distribution have been proposed. The algorithm by \cite{Kent18} is particularly attractive because it uses an acceptance/rejection sampler based on the angular central Gaussian (ACG) distribution as an envelope and is therefore straightforward to implement. We use this method as a baseline against which we compare our slice samplers as well as RWMH and HMC. We run the geodesic slice samplers on the sphere (geoSSS) as well as RWMH and HMC on two high-dimensional Bingham targets with $\dim=10$ and $\dim=50$, respectively. For each target and sampling algorithm, we simulate the consecutive Markov chain realizations $x_1,\dots,x_{N}$ with $N=10^5$ starting from the mode $u_{\dim}$. To explore the variability of all MCMC algorithms, we run 10 repetitions. The step-size parameter of the RWMH and HMC algorithm is tuned as described in Appendix \ref{Sec: Step-size tuning} to achieve a reasonable average acceptance rate.

\begin{figure}[h]
	\centering{\includegraphics[width=0.8\textwidth]{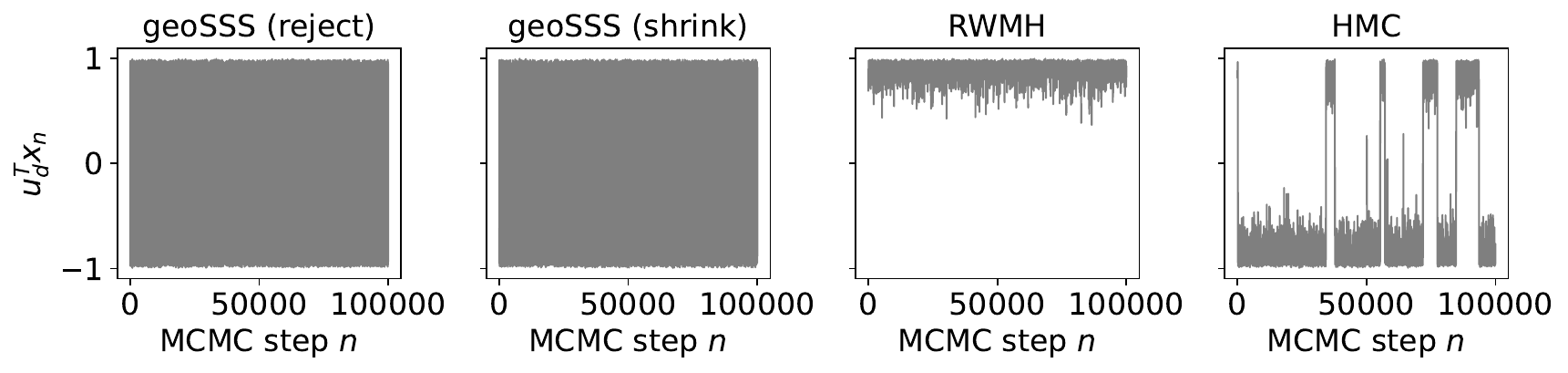}}
	\caption{Traces of approximate samples from a Bingham distribution with $\dim=10$ and $\kappa_{\dim} = 30$ projected onto one of both modes $u_{\dim}$. }\label{fig:bing1-traces}
\end{figure}
The first Bingham target has $\dim=10$ and $\kappa_{\dim} = 30$. In Figure \ref{fig:bing1-traces}, we show traces of approximate Bingham samples projected onto the first mode, that is $u_{\dim}^T x$, which varies between $1$ and $-1$ and is expected to peak somewhat below these extreme values that correspond to both modes. As is evident from the trace plots, both variants of geodesic slice sampling on the sphere find both modes of the Bingham distribution and rapidly jump between them. RWMH, on the other hand, does not escape from the mode in which the chain was started and therefore only finds a single mode. HMC also finds the second mode, but only jumps occasionally between the two modes.

\begin{figure}[h]
	\centering{
		\includegraphics[width=0.8\textwidth]{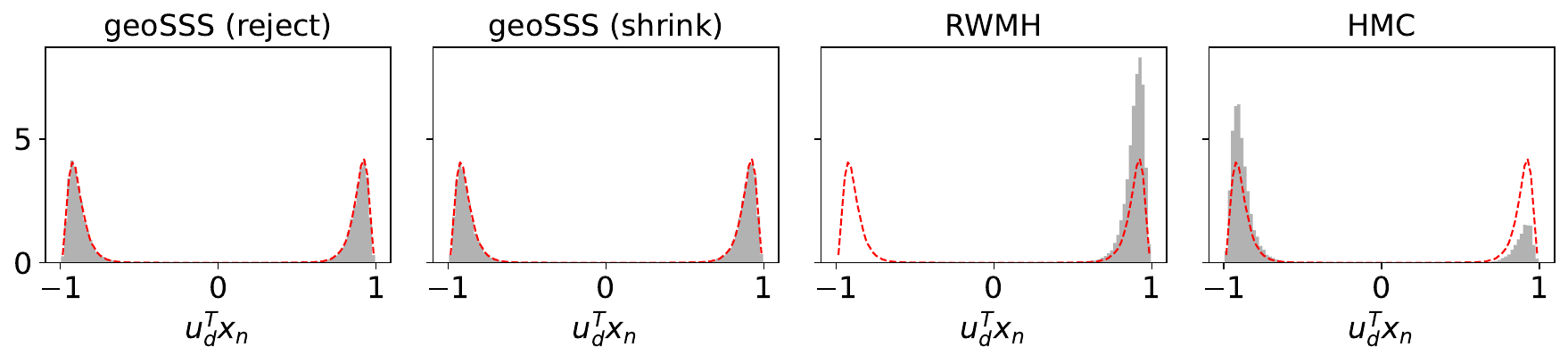}
	}
	\caption{Histograms of approximate Bingham samples with $\dim=10$ and $\kappa_{\dim} = 30$ projected on the first mode obtained with each MCMC method are shown in gray. The red dashed line indicates the baseline obtained with the acceptance/rejection sampler of \cite{Kent18}.}
	\label{fig:bing1-hist}
\end{figure}
The insufficient sampling by RWMH and HMC results in an incorrect exploration of the modes, which should be populated equally. This is illustrated in Figure \ref{fig:bing1-hist} which also shows the distribution of $u_{\dim}^Tx$ obtained by the aforementioned acceptance/rejection method \citep[see][]{Kent18}. The histograms obtained with the geodesic slice samplers closely match the baseline, whereas RWMH completely misses the second mode and HMC misrepresents the probability mass under the modes.

\begin{figure}[h]
	\centering{
		\includegraphics[width=0.8\textwidth]{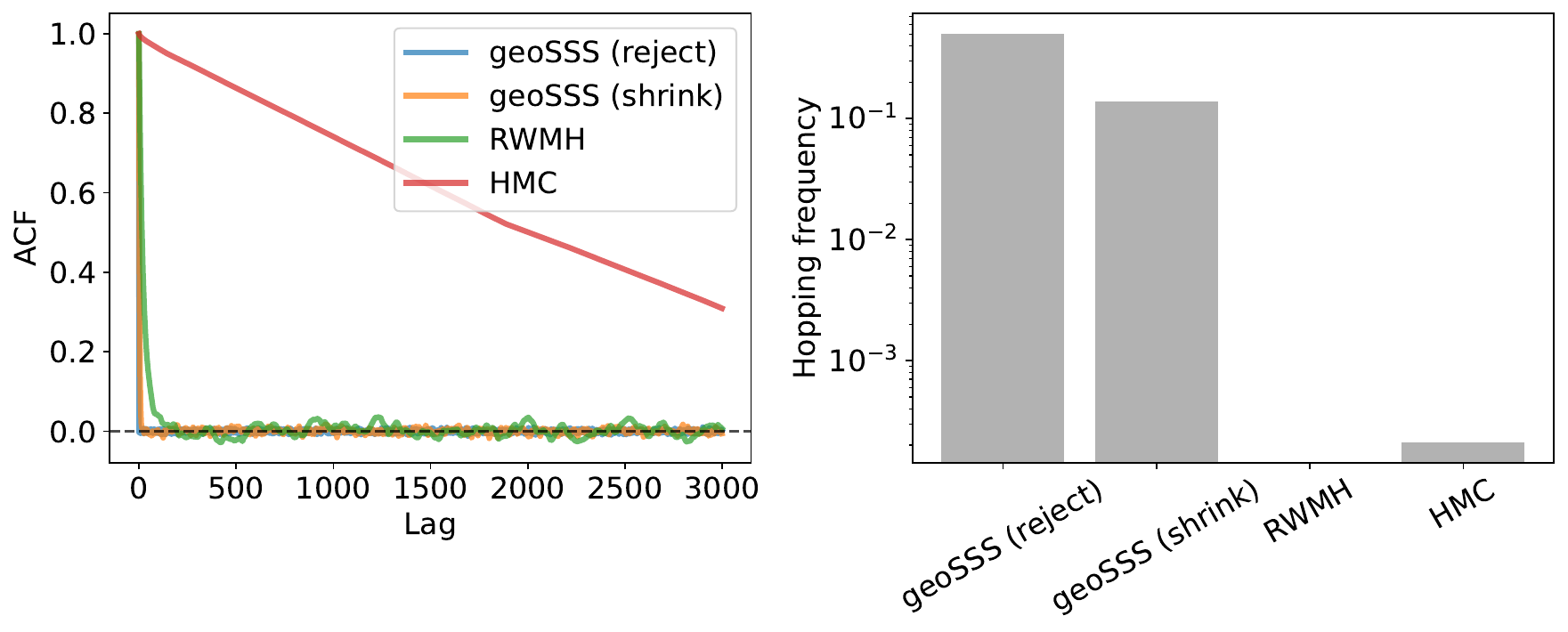}
	}
	\caption{Left: Autocorrelation analysis of $x\mapsto u_{\dim}^Tx$ with respect to approximate samples from a Bingham distribution with $\dim=10$ and $\kappa_{\dim} = 30$. Right: Estimated hopping frequency between both modes of the Bingham distribution. Note that the hopping frequency is shown on a logarithmic axis. }\label{fig:bing1-acf}
\end{figure}

Figure \ref{fig:bing1-acf} shows the autocorrelation function (ACF) of $u_{\dim}^T x$ for the four MCMC samplers. We see a rapid decorrelation in case of both geodesic slice samplers; in fact, the ACF of rejection-based geoSSS drops to zero after a single step. Samples generated with RWMH decorrelate after roughly $200$ MCMC steps, but the faster decorrelation (in comparison to HMC) is due to the fact that RWMH only explores a single mode. HMC finds both modes but shows a very slowly decaying ACF, because jumps between the two modes only occur very rarely. This is also reflected in the effective sample size (ESS). Due to the immediate decorrelation of samples generated with geoSSS using a rejection strategy, the \emph{relative ESS}\footnote{The relative ESS is just the ESS divided by the total number of performed MCMC steps, in our experiments $10^5$.} is estimated to be 99.73\,\%{}. The shrinkage-based geoSSS
obtains a relative ESS of 15.2\,\%{}, whereas RMWH and HMC achieve only very low relative ESS: 0.004\,\%{} and 0.01\,\%{}, respectively.

To quantify how rapidly the MCMC samplers mix between the two modes of the Bingham distribution, we estimated a {\em hopping frequency}, which we define as the average number of times the Markov chain jumps from one mode to the other, given by
\begin{equation}\label{eq:hopping}
	\frac{1}{N-1} \sum_{n=1}^{N-1} \llbracket \text{sign}(u_d^Tx_{n+1}) \not= \text{sign}(u_d^Tx_{n}) \rrbracket,
\end{equation}
where $x_n$ is the realization of the $n$-th Markov chain sample, $N$ the total number of steps and $\llbracket \cdot \rrbracket$ denotes the Iverson bracket.\footnote{For proposition $S$ it holds that $\llbracket S \rrbracket = 1$ if $S$ is a true and $\llbracket S \rrbracket = 0$ otherwise.}

Figure \ref{fig:bing1-acf} shows the estimated hopping frequencies. As expected, the rejection-based geoSSS shows the highest number of oscillations between both modes with approximately 50\%{} hopping frequency, whereas the shrinkage-based geoSSS tends to jump only every seventh step to the other mode. This behavior is expected, because the geodesic level set always contains both modes with equal probability. Therefore, the rejection sampling strategy finds each mode with equal probability, independent of what the current state of the Markov chain is. The shrinkage-based approach has a higher chance to stay in the vicinity of the current state.

\begin{figure}[htbp]
	\centering{
		\includegraphics[width=0.8\textwidth]{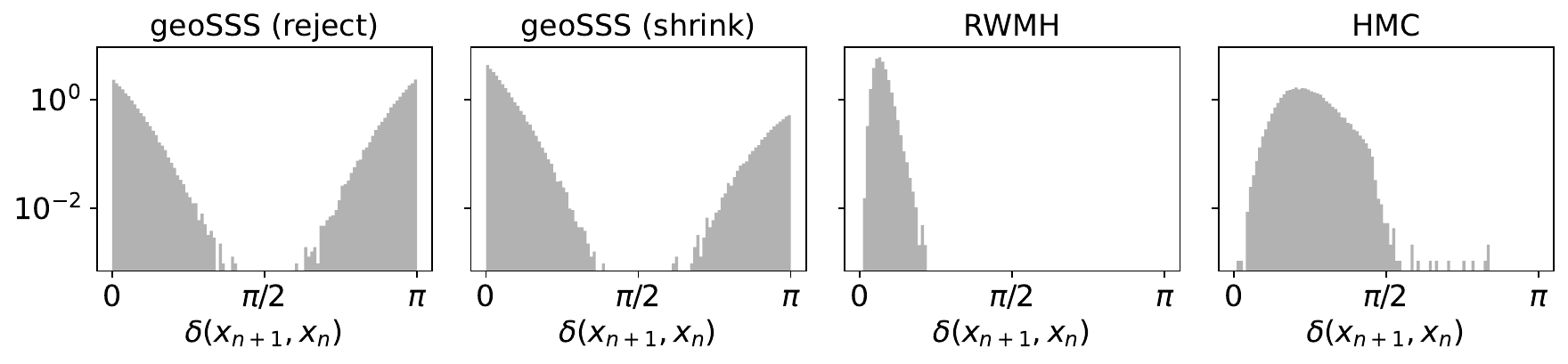}
	}
	\caption{Geodesic distance (log-scale) between successive approximate samples for a Bingham distribution with $\dim=10$ and $\kappa_d=30$.}\label{fig:bing1-dist}
\end{figure}
Another quality measure is the geodesic or great circle distance between successive samples given as
\[
\delta(x_{n+1}, x_n) := \arccos(x_{n+1}^T x_n)\, .
\]
An efficient MCMC algorithm should explore the sphere rapidly by making large leaps from one sample to the next. Again, we see in Fig.~\ref{fig:bing1-dist} a superior performance of geoSSS. The RWMH algorithm achieves only small jumps. As expected, HMC moves more rapidly over the sphere compared to RWMH, but it still cannot compete with the geodesic slice samplers. 

We also run similar tests on a more challenging Bingham target with $\dim=50$ and $\kappa_{\dim} = 300$, i.e., the dimension of the sample space is much larger and the distribution is more concentrated. We observe the same trends as before. Supplementary Figure \ref{fig:bing2-hist} shows the distribution of samples projected onto the first mode. Now, both RWMH and HMC are stuck in the first mode and fail to find the second mode, whereas the geodesic slice samplers represent both modes accurately. The ACF of HMC outperforms RWMH as expected (see Supplementary Fig. \ref{fig:bing2-acf}), but still the geodesic slice samplers show a faster decorrelation than HMC resulting in a higher effective sample size. Since no jumps occur during the entire run of RWMH and HMC, their hopping frequencies are estimated to be smaller than $10^{-5}$, whereas geoSSS still achieves an acceptable jump rate (see Appendix Fig. \ref{fig:bing2-acf}). The rejection-based geoSSS clearly outperforms the shrinkage strategy on this target by achieving a higher hopping rate and as a consequence also a more favorable distribution of step sizes $\delta(x_{n+1}, x_{n})$ (see Appendix Fig. \ref{fig:bing2-dist}).
}

\end{document}